\newdimen\proofrulebreadth \proofrulebreadth=.05em
\newdimen\proofdotseparation \proofdotseparation=1.25ex
\newdimen\proofrulebaseline \proofrulebaseline=2ex
\newcount\proofdotnumber \proofdotnumber=3
\let\then\relax
\def\hfi{\hskip0pt plus.0001fil}
\mathchardef\squigto="3A3B
\newif\ifinsideprooftree\insideprooftreefalse
\newif\ifonleftofproofrule\onleftofproofrulefalse
\newif\ifproofdots\proofdotsfalse
\newif\ifdoubleproof\doubleprooffalse
\let\wereinproofbit\relax
\newdimen\shortenproofleft
\newdimen\shortenproofright
\newdimen\proofbelowshift
\newbox\proofabove
\newbox\proofbelow
\newbox\proofrulename
\def\shiftproofbelow{\let\next\relax\afterassignment\setshiftproofbelow\dimen0 }
\def\shiftproofbelowneg{\def\next{\multiply\dimen0 by-1 }%
\afterassignment\setshiftproofbelow\dimen0 }
\def\setshiftproofbelow{\next\proofbelowshift=\dimen0 }
\def\setproofrulebreadth{\proofrulebreadth}

\def\prooftree{
\ifnum  \lastpenalty=1
\then   \unpenalty
\else   \onleftofproofrulefalse
\fi
\ifonleftofproofrule
\else   \ifinsideprooftree
        \then   \hskip.5em plus1fil
        \fi
\fi
\bgroup
\setbox\proofbelow=\hbox{}\setbox\proofrulename=\hbox{}%
\let\justifies\proofover\let\leadsto\proofoverdots\let\Justifies\proofoverdbl
\let\using\proofusing\let\[\prooftree
\ifinsideprooftree\let\]\endprooftree\fi
\proofdotsfalse\doubleprooffalse
\let\thickness\setproofrulebreadth
\let\shiftright\shiftproofbelow \let\shift\shiftproofbelow
\let\shiftleft\shiftproofbelowneg
\let\ifwasinsideprooftree\ifinsideprooftree
\insideprooftreetrue
\setbox\proofabove=\hbox\bgroup$\displaystyle 
\let\wereinproofbit\prooftree
\shortenproofleft=0pt \shortenproofright=0pt \proofbelowshift=0pt
\onleftofproofruletrue\penalty1
}

\def\eproofbit{
\ifx    \wereinproofbit\prooftree
\then   \ifcase \lastpenalty
        \then   \shortenproofright=0pt  
        \or     \unpenalty\hfil         
        \or     \unpenalty\unskip       
        \else   \shortenproofright=0pt  
        \fi
\fi
\global\dimen0=\shortenproofleft
\global\dimen1=\shortenproofright
\global\dimen2=\proofrulebreadth
\global\dimen3=\proofbelowshift
\global\dimen4=\proofdotseparation
\global\count255=\proofdotnumber
$\egroup  
\shortenproofleft=\dimen0
\shortenproofright=\dimen1
\proofrulebreadth=\dimen2
\proofbelowshift=\dimen3
\proofdotseparation=\dimen4
\proofdotnumber=\count255
}

\def\proofover{
\eproofbit 
\setbox\proofbelow=\hbox\bgroup 
\let\wereinproofbit\proofover
$\displaystyle
}%
\def\proofoverdbl{
\eproofbit 
\doubleprooftrue
\setbox\proofbelow=\hbox\bgroup 
\let\wereinproofbit\proofoverdbl
$\displaystyle
}%
\def\proofoverdots{
\eproofbit 
\proofdotstrue
\setbox\proofbelow=\hbox\bgroup 
\let\wereinproofbit\proofoverdots
$\displaystyle
}%
\def\proofusing{
\eproofbit 
\setbox\proofrulename=\hbox\bgroup 
\let\wereinproofbit\proofusing
\kern0.3em$
}

\def\endprooftree{
\eproofbit 
  \dimen5 =0pt
\dimen0=\wd\proofabove \advance\dimen0-\shortenproofleft
\advance\dimen0-\shortenproofright
\dimen1=.5\dimen0 \advance\dimen1-.5\wd\proofbelow
\dimen4=\dimen1
\advance\dimen1\proofbelowshift \advance\dimen4-\proofbelowshift
\ifdim  \dimen1<0pt
\then   \advance\shortenproofleft\dimen1
        \advance\dimen0-\dimen1
        \dimen1=0pt
        \ifdim  \shortenproofleft<0pt
        \then   \setbox\proofabove=\hbox{%
                        \kern-\shortenproofleft\unhbox\proofabove}%
                \shortenproofleft=0pt
        \fi
\fi
\ifdim  \dimen4<0pt
\then   \advance\shortenproofright\dimen4
        \advance\dimen0-\dimen4
        \dimen4=0pt
\fi
\ifdim  \shortenproofright<\wd\proofrulename
\then   \shortenproofright=\wd\proofrulename
\fi
\dimen2=\shortenproofleft \advance\dimen2 by\dimen1
\dimen3=\shortenproofright\advance\dimen3 by\dimen4
\ifproofdots
\then
        \dimen6=\shortenproofleft \advance\dimen6 .5\dimen0
        \setbox1=\vbox to\proofdotseparation{\vss\hbox{$\cdot$}\vss}%
        \setbox0=\hbox{%
                \advance\dimen6-.5\wd1
                \kern\dimen6
                $\vcenter to\proofdotnumber\proofdotseparation
                        {\leaders\box1\vfill}$%
                \unhbox\proofrulename}%
\else   \dimen6=\fontdimen22\the\textfont2 
        \dimen7=\dimen6
        \advance\dimen6by.5\proofrulebreadth
        \advance\dimen7by-.5\proofrulebreadth
        \setbox0=\hbox{%
                \kern\shortenproofleft
                \ifdoubleproof
                \then   \hbox to\dimen0{%
                        $\mathsurround0pt\mathord=\mkern-6mu%
                        \cleaders\hbox{$\mkern-2mu=\mkern-2mu$}\hfill
                        \mkern-6mu\mathord=$}%
                \else   \vrule height\dimen6 depth-\dimen7 width\dimen0
                \fi
                \unhbox\proofrulename}%
        \ht0=\dimen6 \dp0=-\dimen7
\fi
\let\doll\relax
\ifwasinsideprooftree
\then   \let\VBOX\vbox
\else   \ifmmode\else$\let\doll=$\fi
        \let\VBOX\vcenter
\fi
\VBOX   {\baselineskip\proofrulebaseline \lineskip.2ex
        \expandafter\lineskiplimit\ifproofdots0ex\else-0.6ex\fi
        \hbox   spread\dimen5   {\hfi\unhbox\proofabove\hfi}%
        \hbox{\box0}%
        \hbox   {\kern\dimen2 \box\proofbelow}}\doll%
\global\dimen2=\dimen2
\global\dimen3=\dimen3
\egroup 
\ifonleftofproofrule
\then   \shortenproofleft=\dimen2
\fi
\shortenproofright=\dimen3
\onleftofproofrulefalse
\ifinsideprooftree
\then   \hskip.5em plus 1fil \penalty2
\fi
}

\documentstyle{article}
\begin{document}
\title{Double-Negation Elimination in Some Propositional Logics}
\author{{\em Michael Beeson}\\
San Jose State University\\
Math \& Computer Science\\
San Jose, CA 95192 \\
\and
{\em Robert Veroff}\\
University of New Mexico\\
Department of Computer Science\\
Albuquerque, NM 87131 \\
\and
{\em Larry Wos}\\
 Mathematics and Computer Science Division\\
           Argonne National Laboratory\\
           Argonne, IL  60439-4801\\
            }
\maketitle
\def\bul{$\bullet$\ }
\def\implies{\rightarrow}
\def\imp{\rightarrow}
\def\all{\forall}
\def\seq{\Rightarrow}
\medskip
\newtheorem{theorem}{Theorem}
\newtheorem{lemma}{Lemma}
\newtheorem{corollary}{Corollary}

\section*{Abstract}

This article answers two questions (posed in the literature), each concerning
the guaranteed existence of proofs {\em free of double negation}.  A proof
is free of double negation if none of its deduced steps contains a term of
the form $n(n(t))$ for some term $t$, where $n$ denotes negation.
The first question asks for conditions on the hypotheses that, if
satisfied, guarantee the existence of a double-negation-free proof
when the conclusion is free of double negation.  The second question asks
about the existence of an axiom system for classical propositional calculus
whose use, for theorems with a conclusion free of double negation, guarantees
the existence of a double-negation-free proof.  After giving conditions that
answer the first question, we answer the second question by focusing on the
{\L}ukasiewicz three-axiom system.  We then extend our studies to
infinite-valued sentential calculus and to intuitionistic logic and
generalize the notion of being double-negation free. The double-negation
proofs of interest rely exclusively on the inference rule condensed
detachment, a rule that combines modus ponens with an appropriately general
rule of substitution.  The automated reasoning program OTTER played an
indispensable role in this study.

\section{Origin of the Study}
 
This article features the culmination of a study whose origin rests equally with two questions, the first posed in {\em Studia Logica} \cite{fitelson} and the second (motivated by the first) posed in the {\em Journal of Automated Reasoning} \cite{ulrich}.
Both questions focus on {\em double-negation-free proofs}, proofs none of whose deduced steps contain a formula of the form $n(n(t))$ for some term $t$ with the function $n$ denoting negation.
For example, where $i$ denotes implication, the presence of the formula $i(i(n(x),x),x)$ as a deduced step does not preclude a proof from being double-negation free, whereas the presence of the formula $i(n(n(x)),x)$ does.
Note the distinction between deduced steps and axioms; in particular, use of the Frege system for two-valued sentential calculus, which contains two axioms in which double negation occurs, guarantees the existence of double-negation-free proofs, as we show in Section 5.
  
The sought-after double-negation-free proofs of interest here rely solely on the inference rule condensed detachment~\cite{prior-play}, a rule that combines modus ponens with an appropriately general rule of substitution.
Formally, condensed detachment considers two formulas, $i(A,B)$ (the major premiss) and $C$ (the minor premiss), that are tacitly assumed to have no variables in common, and, if $C$ unifies with $A$, yields the formula $D$, where $D$ is obtained by applying to $B$ a most general unifier of $C$ and $A$.
 
In \cite{fitelson}, the following question is asked.
Where $P$ and $Q$ may each be collections of formulas, if {\bf T} is a theorem asserting the deducibility of $Q$ from $P$ such that $Q$ is free of double negation, what conditions guarantee that there exists a proof relying solely on condensed detachment all of whose deduced steps are free of double negation?
Then, in \cite{ulrich}, Dolph Ulrich asks about the existence of an axiom system for two-valued sentential (or classical propositional) calculus such that, for each double-negation-free formula $Q$ provable from the axiom system, there exists a double-negation-free proof of $Q$.
  
Although perhaps not obvious, the nature of the axioms chosen for the study of some area of logic or mathematics can have a marked impact on the nature of the proofs derived from them.
As a most enlightening illustration of this relation and indeed pertinent to the two cited questions (each of which we answer in this article), we turn to an example given by Ulrich that builds on a result of C. A. Meredith.
In the early 1950s, Meredith found the following 21-letter single axiom for two-valued logic.
$$
i(i(i(i(i(x,y),i(n(z),n(u))),z),v),i(i(v,x),i(u,x)))
$$

Consider the following system with condensed detachment as the sole rule of inference and the four double-negation-free classical theses (of two-valued logic) as axioms. (The notation here is taken from Ulrich \cite{ulrich} and should {\em not} be confused with that used for infinite-valued logic discussed in Section 7.)

\begin{eqnarray*}
\mbox{A1}&\hskip 0.15in &i(x,x)\\
\mbox{A2}&&i(i(x,x),i(n(x),i(n(x),n(x))))\\
\mbox{A3}&&i(i(x,i(x,x)),i(n(x),i(n(x),i(n(x),n(x)))))\\
\mbox{A4}&&i(i(x,i(x,i(x,x))),i(i(i(i(i(y,z),i(n(u),(v))),u),w),i(i(w,y),i(v,y))))
\end{eqnarray*}

One can readily verify that axiom A1 and the antecedent (left-hand argument) of A2 are unifiable but that no other axiom is unifiable with the antecedent of any axiom.
In other words, no conclusion can be drawn (with condensed detachment) other than by considering A1 and A2.
Therefore, the first step of any proof in this system can only be

\vspace{.15in}
\noindent
5 $\qquad i(n(x),i(n(x),n(x))).$

\vspace{.15in}
Similarly, the only new path of reasoning now available is that of 5 with the antecedent of A3.
Therefore, the next step in any proof in this system can only be

\vspace{.15in}
\noindent
6 $\qquad i(n(n(x)),i(n(n(x)),i(n(n(x)),n(n(x))))).$

\vspace{.15in}
Of course, 6 and the antecedent of A4 are unifiable, and we may obtain

\vspace{.15in}
\noindent
7 $\qquad i(i(i(i(i(x,y),i(n(z),n(u))),z),v),i(i(v,x),i(u,x))).$

\vspace{.15in}

But, since 7 is Meredith's single axiom for two-valued sentential calculus, we may then deduce all other theorems of classical sentential logic.
The given four-axiom system does, therefore, provide a complete axiomatization for classical $i-n$ (two-valued logic); but no proof of any classical theses except A1--A4 and 5 can be given that does not include at least formula 6, in which $n(n(x))$ (double negation) appears four times.
  
Thus one sees that some axiom systems have so much control over proofs derived from them that double negation is inescapable.
As for the Meredith single axiom (derived from the Ulrich example), what is its status with regard to guaranteed double-negation-free proofs of theorems that themselves are free of double negation?
Of a sharply different flavor, what is the status in this regard of the Frege axiom system in view of the fact that two of its members each contain a double negation, $i(n(n(x)),x)$ and $i(x,n(n(x)))$?
The Frege axiom system consists of the following six axioms.
\begin{eqnarray*}
\mbox{ } & i(x,i(y,x)).\\
\mbox{ } & i(x,n(n(x))).\\
\mbox{ } & i(n(n(x)),x).\\
\mbox{ } & i(i(x,i(y,z)),i(i(x,y),i(x,z))).\\
\mbox{ } & i(i(x,y),i(n(y),n(x))).\\
\mbox{ } & i(i(x,i(y,z)),i(y,i(x,z))).
\end{eqnarray*}
These questions are also answered in this article as we complete our treatment of two-valued sentential calculus by giving conditions that, if satisfied by the axioms, guarantee the existence of a double-negation-free proof for each theorem that itself is double-negation free.
  
The study of this logical property of obviating the need for double negation demands its examination in other areas of logic and demands a natural extension.
Therefore, we investigate this property in the context of infinite-valued sentential calculus and intuitionistic logic, and we present an extension of the property that focuses on theorems in which double negation appears.
  
Our interest in double-negation avoidance can be traced directly to our successes in using William McCune's automated reasoning program OTTER~\cite{otter}.
In particular, a large number of proofs were obtained with that program by applying a strategy that instructs OTTER to avoid retention of any deduced conclusion if it contains a double-negation term.
Use of this strategy sharply increased the likelihood of success.
Because the literature strongly suggests that reliance on double negation is unavoidable, and because our completed proofs suggested the contrary, the questions that are central to this article were studied.

\section{The Interplay of Axioms and Proof}
 
Once posed, the question of double-negation avoidance seems quite natural,
meshing well with other concerns for proof properties as expressed by logicians.
For example, length of proof was studied by Meredith and Prior, by Thomas, and by others; size of proof (total number
of symbols) is of interest to Ulrich; and the dispensing with thought-to-be-key lemmas is almost always of general interest.
 
More familiar to many are similar concerns for the axioms of a theory.
Indeed, in logic, merited emphasis is placed on the nature and properties of various axiom systems:
the number of members, the length (individually and collectively),
the number of distinct letters (variables), the total number of occurrences
of various function symbols, and other measures of ``simplicity''.
To mention but one of many examples, in the mid-1930s J. {\L}ukasiewicz  discovered a 23-letter single axiom for
two-valued sentential (or classical propositional) calculus.
As cited in Section 1, almost two decades later Meredith found a 21-letter single axiom.
Whether a still shorter single axiom for this area of logic exists is currently unknown.

To date, studies have focused on the properties of proof or the properties of axiom systems; we know of little work that connects the two directly.
Here we study such a direct connection when we show that a double-negation-free proof must always exist when the axioms satisfy certain properties.
In other words, we focus on a term-structure property of proof in its relation to a set of axiom-system properties.
One might naturally wonder about other theorems that provide a direct connection of some proof property with the properties of the axioms under consideration.

Double-negation-free proofs, in addition to their aesthetic appeal and their interest
from a logical viewpoint, are relevant to the work of Hilbert.
Indeed, although it was unknown until recently \cite{thiele}, Hilbert offered a twenty-fourth
problem that was not included in the famous
list of twenty-three seminal problems that he presented in Paris at the beginning of the twentieth century.
This twenty-fourth problem focuses on the finding of simpler proofs and its value.
Hilbert did not include the problem in his Paris talk apparently because
of the difficulty of defining ``simpler'' precisely.

Ceteris paribus, the avoidance of some type of term can make a proof simpler, as
is the case when a proof is free of doubly
negated subformulas.  This paper, in the spirit of Hilbert's twenty-fourth
problem, studies this specific
form of simplicity, seeking (as noted) general sufficient conditions for an axiom system
of propositional logic L that
guarantees that doubly negated formulas that do not occur in the theorem
are not needed in the proof.

\section{Formalism}

Although propositional calculus is one of the oldest areas of logic, not all
of its
mysteries have been unlocked.   The existence of truth tables and other
decision procedures for propositional logic notwithstanding, it is by no
means trivial to
prove, for example, that a given 23-symbol formula is in fact a single
axiom.  Truth tables
and decision procedures can be used to determine whether a given formula is a
tautology  or
to construct a proof of a given formula from certain axioms and rules, but
generally they
are not helpful in finding proofs of known axioms from other formulas
(which is
what one must do to verify that a formula is a single axiom).  The search
for such proofs has recently become
a test bed in automated deduction.  Not only do the theorems we prove here about double-negation elimination have
an intrinsic, aesthetic appeal in that they show the possibility of
simplifying proofs,  but they
also are of interest because they
justify in the vast majority of cases a shortcut in automated proof-search methods, namely, the automatic discarding of double negations.

We shall work with logics formulated by using only the two connectives
implication and negation.
Several notations are in use for propositional logic that we mention
before continuing.  First,
one can use infix $\implies$ for implication and prefix $\neg$ for negation.
For example, we could
write $x \implies (\neg x \implies y)$.  Closely related, many papers on propositional logic
use Polish notation, in which
${\bf C}$ is used for implication (conditional) and ${\bf N}$ for negation.
The same formula would then
be rendered as ${\bf C} x {\bf C} {\bf N} xy$.  Finally, the notation that
is appropriate when using
OTTER is prefix, with parentheses. We use $i(x,y)$ for implication and
$n(x)$ for negation; therefore, the
example formula would be $i(x,i(n(x),y))$.  In this paper we use this last
notation exclusively.
It permits us to cut and paste machine-produced proofs, eliminating errors
of transcription.   We make use
of the theorem-proving program OTTER~\cite{otter} to produce proofs in
various propositional logics, proofs
we use to verify that those logics satisfy the hypotheses of our general
theorems on double-negation elimination.

Let L be {\L}ukasiewicz's  formulation of propositional calculus in terms of
implication and negation,
denoted by $i$ and $n$, as
given on page 221 of \cite{wos}.  {\L}ukasiewicz provided the following axiomatization of L.
\begin{eqnarray*}
\mbox{L1}& \qquad i(i(x,y),i(i(y,z),i(x,z))) \\
\mbox{L2}& \qquad i(i(n(x),x),x) \\
\mbox{L3}& \qquad i(x,i(n(x),y))
\end{eqnarray*}
The inference rule frequently used in logic is known as condensed
detachment.
This rule (which is the only inference rule to be used in the sought-after double-negation-free proofs) combines substitution and modus ponens.
Specifically, given a major
premiss $i(p,q)$  and a minor premiss $p$, the conclusion of modus ponens is
$q$.
The substitution rule permits the deduction of $p\sigma$ from $p$, where
$\sigma$
is any substitution of terms for variables.  Condensed detachment has premisses $i(p,q)$ and $r$
and attempts to unify $p$ and $r$---that is, seeks a substitution $\sigma$ that
makes
$p\sigma = r\sigma$.  If successful,  provided $\sigma$ is the most
general such
substitution,  the conclusion of condensed detachment is $q\sigma$.  This
inference rule requires renaming of variables in the premisses before the
attempted unification to
avoid unintended clashes of variables.%
\footnote{In the absence of the substitution rule,  any alphabetic variant
of an axiom
is also accepted as an axiom.  An ``alphabetic'' variant of $A$ is a formula
$A\sigma$,
where the substitution $\sigma$ is one-to-one and merely renames the
variables. A technicality
arises as to whether it is permitted, required, or forbidden to rename the
variables of the premisses before
applying condensed detachment.
The definition on p. 212 of \cite{wos} does not explicitly mention renaming and, read
literally, would not
allow it, but the implementation in OTTER requires it, and
\cite{ulrich} explicitly permits it.  If it is not permitted, then by
renaming variables in the
entire proof of the premiss, we obtain a  proof of the renamed premiss,
using alphabetic variants of the axioms,
so the same formulas will be provable in either case.  Similarly, renaming
of variables in conclusions is allowed.  Technically,
we could wait until the conclusions are used before renaming them, but in
practice, OTTER renames variables in
each conclusion as it is derived.
}

A double negation is a formula $n(n(t))$, where $t$ is any term.
A formula $A$ {\em contains a double negation} if it has a not-necessarily-proper subformula that
is a double negation.
A derivation contains a double negation if one of its deduced formulas contains a
double negation.
Suppose that the formula $A$ contains no double negations and is derivable
in L.
Then (central to this paper) does $A$ have a derivation in L that contains no double negation?  We
answer
this question in the affirmative (and thus answer the cited Ulrich question),
not only for {\L}ukasiewicz's system L1--L3,
but also for
other axiomatizations of classical (two-valued) propositional logic, as well as other systems of
logic 
such as infinite-valued logic.

\section{Condensed Detachment}
We remind the reader that the systems of primary
interest in this paper
use condensed detachment as their {\em sole} rule of
inference. For example,
if $\alpha$ is a complicated formula, and we wish to
deduce $i(\alpha,\alpha)$,
it would not be acceptable to first deduce $i(x,x)$ and
then substitute $\alpha$ for $x$.
Rather, it would be necessary to give a (longer) direct derivation
of $i(\alpha,\alpha)$, relying solely on applications of condensed detachment.

We shall show in this section that our theorem about
the eliminability of
double negation holds for L1--L3 with condensed
detachment if and only if
it holds for L1--L3 with modus ponens and substitution.
Similar results are in \cite{cd1,cd3}, but for other
systems \cite{cd1} treats the implicational fragment, while
we allow negation, and \cite{cd3} treats relevance logic.
The following three
formulas will play an
important role.

\begin{eqnarray*}
\mbox{ D1} & \qquad  i(x,x) \\
\mbox{ D2} & \qquad i(i(x,x),i(n(x),n(x)))  \\
\mbox{ D3} & \qquad
i(i(x,x),i(i(y,y),i(i(x,y),i(x,y))))
\end{eqnarray*}

\begin{lemma}  \label{lemma:ialphaalpha}
Suppose L is any system of propositional logic with
condensed detachment as the sole inference rule,
and suppose that there are proofs
of D1--D3 in L.
Then every formula of the form $i(\alpha,\alpha)$ is
provable from
L by condensed detachment.  Furthermore, if there are
double-negation-free proofs of D1--D3 in L, then $i(\alpha,\alpha)$ 
is provable without using double
negations except those occurring as subformulas of
$\alpha$.
\end{lemma}

\noindent{\em Proof\/}.
We prove by induction on the complexity of the
propositional formula $\alpha$ that
for each $\alpha$,  the formula $i(\alpha,\alpha)$ is
provable in L by condensed
detachment.  The base case, when $\alpha$ is a
proposition letter,  follows by
replacing $x$ by $\alpha$ in the proof of $i(x,x)$.
Any line of the proof that is an axiom
becomes an alphabetic variant of that axiom, which is
still considered an axiom.  Actually, in view of the
convention that
renaming variables in the conclusion is allowed, it
would be enough just to replace $x$ by
$\alpha$ in the last line of the proof.
If we have a proof of $i(\beta,\beta)$,
then we can apply condensed detachment and D2 to get a
proof of
$i(n(\beta),n(\beta))$.  This could introduce a double
negation if $\beta$ is already
a negation, but in that case it is a double negation
that already occurs in $\alpha = n(\beta)$,
and so is allowed. Similarly, if we have proofs of
$i(\alpha,\alpha)$ and
$i(\beta,\beta)$, we can apply condensed detachment to
D3 and
get a proof of $i(i(\alpha,\beta),i(\alpha,\beta))$.
That completes the proof of the lemma.

\begin{lemma}
\label{lemma:six}
If $A$ is an instance of $C$, then the result of
applying condensed detachment to
$i(A,B)$ and $C$ is $B$ (or an alphabetic variant of
$B$).
\end{lemma}

\noindent{\em Proof\/}:  Rename variables in $C$ if
necessary so that
$C$ and $A$ have no variables in common.  Let $\sigma$
be a most general
unifier of $A$ and $C$.  Then the result of applying
condensed detachment
to $i(A,B)$ and $C$ is $B\sigma$.

Let $\tau$ be a most general substitution such that
$C\tau = A$; since
$A$ is assumed to be an instance of $C$, such a $\tau$
exists.  Since
the variables of $C$ do not occur in $A$ or $B$, $B\tau
= B$ and
$A\tau = A$.  Then $C\tau = A = A\tau$, so $\tau =
\sigma \rho$ for
some substitution $\rho$.  Then $B = B\tau =
B\sigma\rho$.  Thus
$\sigma\rho$ is the identity on $B$.  Hence
$\sigma\rho$ is the
identity on each variable occurring in $B$.  Hence
$\sigma$ and $\rho$
do nothing but (possibly) rename variables.  Hence
$B\sigma$,
which is the result of this application of condensed
detachment, is
$B$ or an alphabetic variant of $B$.  That completes
the proof of the lemma.

\begin{lemma} \label{lemma:axsub} Suppose L is a logic
proving D1--D3 by condensed detachment.  
Then each substitution instance $\alpha$ of
an axiom of L is provable by
condensed detachment.  Furthermore, if L proves
D1--D3 by condensed detachment without using double negations,
then $\alpha$ is also provable without using double negations,
except those double negations
occurring as subformulas of $\alpha$, if any.
\end{lemma}

\noindent{\em Proof\/}. Let $\alpha$ be a substitution
instance of an axiom $A$.
Renaming the variables in the axiom $A$ if necessary,
we may assume that
the variables occurring in $A$ do not occur in
$\alpha$.
By Lemma \ref{lemma:ialphaalpha}, $i(\alpha,\alpha)$ is
provable by
condensed detachment, without using any double
negations except possibly those
already occurring in $\alpha$.   By Lemma
\ref{lemma:six}, the result of
applying condensed detachment to $i(\alpha,\alpha)$ and
$A$ is $\alpha$ or an
alphabetic variant $\alpha\sigma$ of $\alpha$.  If it
is not literally $\alpha$, we can rename variables
in the conclusion (or if one prefers to avoid renaming
conclusions,
throughout the entire proof) to create a proof of
$\alpha$.  This completes the proof
of the lemma.

A proof of $B$ in L from assumptions $\Gamma$ is
defined as usual: Lines of the proof
are inferred from previous lines, or are axioms,
or belong to $\Gamma$.
When condensed detachment is used as a rule of
inference, however, we have to distinguish between
(propositional) variables that occur in the axioms and
specific (constant) proposition letters
that occur in assumptions.   For example, if we have
$i(n(n(x)),x)$ as an axiom, then we can
derive any substitution instance of that formula,  but
if we have $i(n(n(a)),a)$ as an assumption,
we cannot use it to derive an instance with some other
formula substituted for $a$.

The following theorem is the easy half of the relation
between condensed-detachment proofs and modus ponens
proofs.
The sense of the theorem is that substitutions can be
pushed back to the axioms.

\begin{theorem}[Pushback theorem]
\label{theorem:pushback}  Let L be a system of
propositional logic, and suppose
L proves $B$ by using condensed detachment, or by using modus
ponens and substitution.
Then there exists a proof of $B$ using modus ponens
from
substitution instances of axioms of L.   Similarly, if L
proves $B$ from assumptions $\Delta$, then there exists a
proof of $B$ using modus ponens from $\Delta$ and
substitution instances of axioms of L.
\end{theorem}

\noindent{\em Remark}.  It would not make sense to
speak of substitution instances of $\Delta$, since
assumptions cannot contain variables,  as explained
above.

\noindent{\em Proof}.  First we prove the theorem for the case when
the given proof uses
modus ponens and substitution.  We proceed by induction
on the length of the given proof of $B$.
If the length is zero, then $B$ is an axiom or
assumption, and there is nothing to prove.
If the last inference is by modus ponens,
say $B$ is inferred from $i(A,B)$ and $A$, then by the
induction hypothesis there exist proofs of these
premisses from substitution instances of axioms, and
adjoining the last inference, we obtain the
desired proof of $B$.

If the last inference is by substitution, say $B =
A\sigma$ is inferred
from $A$, then by the induction hypothesis there exists
a proof $\pi$ of $A$ using modus ponens
only from substitution instances of axioms. Apply the
substitution $\sigma$ to every line of $\pi$;
the result is the desired proof of $B$.   If there are
assumptions,  they are unaffected by $\sigma$
because they do not contain variables.

Now suppose the original proof uses condensed
detachment.
Each condensed-detachment inference can be broken into
two substitutions and an application of modus ponens, so
a condensed-detachment proof gives rise to a modus
ponens and substitution proof, and we can apply
the previous part of the proof.  That completes the
proof.

The following lemma is not actually used in our work but 
is of independent interest.  Condensed detachment is considered 
as an inference rule that combines modus ponens and substitution. 
The following lemma shows that it is reasonable to consider systems
whose only rule of inference is condensed detachment, because such systems
are already closed under the rule of substitution.  This is not obvious 
{\em a priori} since condensed detachment permits only certain special substitutions.

\begin{lemma} \label{lemma:penultimate}  Suppose L is a
logic proving formulas D1--D3 by
condensed detachment
If $A$ is provable in L with condensed detachment and
$\sigma$ is
any substitution, then $A \sigma$ is provable in L by
condensed detachment.   
\end{lemma}

\noindent{\em Proof\/}.  By induction on the length of
the proof $\pi$ of $A$ in L, we
prove that the statement of the lemma is true for all
substitutions $\sigma$.
The base case occurs when $A$ is an axiom, so $A\sigma$
is a substitution instance
of an axiom. By Lemma \ref{lemma:axsub}, $A\sigma$ is
provable in L by condensed detachment.

For the induction step, suppose the last inference of
the given proof $\pi$
has premisses $i(p,q)$ and $r$,
where $\tau$ is the most general unifier of $p$ and
$r$,  and the conclusion is $q\tau = A$.
By the induction hypothesis, we have
condensed-detachment derivations of
$i(p\tau\sigma,q\tau\sigma)$
and of $r\tau\sigma$.  Since $p\tau = r \tau$, also $p
\tau\sigma = r \tau\sigma$. Hence
the inference from $i(p\tau\sigma,q\tau\sigma)$ and $r
\tau \sigma$  to $q \tau \sigma$ is
legal by condensed detachment.  Hence we have a
condensed-detachment proof of $q \tau \sigma =
A\sigma$.
This  completes the proof of the lemma.

\begin{theorem}[D-completeness]
\label{theorem:dcomplete} Suppose L is a logic that
proves
formulas D1--D3.  If L proves $A$ by using modus ponens
and substitution, then
L proves $A$ by using condensed detachment.
\end{theorem}

\noindent{\em Remark\/}.  Note that we cannot track
what happens to double negations in this proof.
The proof does not guarantee that passing from
substitution to condensed detachment will not introduce
new double negations.  Somewhat to our surprise, we do
not need any such result to prove double negation
elimination; indeed, quite the reverse, we shall derive
such a result from double-negation elimination.

\noindent{\em Proof\/}.   By Theorem
\ref{theorem:pushback}, there exists a proof $\pi$ of
$A$
from  substitution instances of axioms,
using modus ponens as the only rule of inference.
By Lemma \ref{lemma:axsub}, there exist
condensed-detachment proofs of these substitution
instances of axioms.  Since modus ponens is a special
case of condensed detachment, if we
string together the condensed-detachment proofs of the
instances of axioms required,
followed by the proof $\pi$, we obtain a condensed-detachment proof of $A$.
That completes the proof of the theorem.

\section{The Main Theorem}

Let L be a system of propositional logic, given by some
axioms and the sole inference
rule of condensed detachment.
Let L* be the system of logic whose axioms are the
closure of (the axioms of) L under
applications of the following syntactic rule:  If $x$
is a proposition letter,
and subterm $n(x)$ appears in a formula $A$, then
construct a new formula by
replacing each occurrence of $x$ in $A$ by $n(x)$ and
cancelling any double
negations that result.   In other words,  we choose a
set $S$ of proposition letters occurring negated in
$A$,
and we replace each occurrence of a variable $x$ in $S$
throughout $A$  by $n(x)$,
cancelling any doubly negated propositions.
The first description of L* calls for replacing all
occurrences of only {\em one} variable;
but if we repeat that operation, we can in effect
replace a subset.

An example will make the definition of L* clear.  If
this procedure is applied to the axiom
$$ i(i(n(x),n(y)),i(y,x)),$$
we obtain the following three new axioms (by replacing
first both $x$ and $y$, then only $y$, then only $x$).
\begin{eqnarray*}
\mbox{A6}& \qquad i(i(x,y),i(n(y),n(x))) \\
\mbox{A7}& \qquad i(i(n(x),y),i(n(y),x))  \\
\mbox{A8}& \qquad i(i(x,n(y)),i(y,n(x)))
\end{eqnarray*}

We say that L admits double-negation elimination if,
whenever L proves a theorem $B$ of the form $P$ implies $Q$, there exists
a proof $S$ of $B$ in L such that any double negations occurring as
subformulas in the deduced steps of $S$ occur as subformulas
of $Q$.\footnote{In this context, a formula $t$ occurs as a subformula
if and only if $t$ or an alphabetic variant of $t$ appears.
For example, if $n(n(i(u,u)))$ occurs in $Q$, then $n(n(i(x,x)))$ would
be permitted in the deduced steps of $S$ but not $n(n(i(x,y)))$
or $n(n(i(z,z),i(z,z)))$.}
In particular, {\em double-negation-free
theorems have double-negation-free proofs} (ignoring the axioms).

Suppose $B$ contains several doubly negated
subformulas.  We wish to consider eliminating
double negations on just {\em some} of those
subformulas.  Let a subset of the doubly negated
formulas in $B$ be selected.  Then let $B^*$ be the
result of erasing double negations on
{\em all occurrences} of the selected subformulas in
$B$.  More precisely,
$B^*$ is obtained from $B$ by replacing all occurrences
of
selected doubly negated subformulas $n(n(q))$ in $B$ by
$q$.
We emphasize that if some
doubly negated subformula occurs more than once in $B$,
one must erase double negations on
all or none of those  occurrences.  
Generally there will be more than one way to select a
set of doubly negated subformulas, so $B^*$ is not
unique.
We say that L admits strong double-negation elimination
if, whenever L proves a theorem $B$,
and $B^*$ is obtained from $B$ as described,  then
there exists a proof of $B^*$ in L, and moreover, there
exists a proof of $B^*$ in L that contains
only doubly negated formulas occurring in $B^*$.

\begin{theorem} \label{theorem:main}  Suppose that in L
there exist double-negation-free proofs of D1--D3,
and double-negation-free proofs of all the axioms of
L*.   Then L admits strong double-negation elimination.
\end{theorem}
\noindent{\em Remark}.  The theorem is also true with
triple negation or quadruple negation, and so forth,
in place of double negation.  For instance, if $B$
contains a triple negation, then it has a proof
containing no double negations not already contained in
$B$.  In particular, it then contains no
triple negations not already contained in $B$, since
every triple negation is a double negation.

\noindent{\em Proof}.  Suppose $B$ is provable in L.
If $B$ contains any double negations,
select arbitrarily a subset of the doubly negated
subformula of $B$, and form $B^*$ by replacing
each occurrence of these formulas $n(n(q))$ by $q$.  Of
course, $B^*$ may still contain double negations;  if
we
are proving only double-negation elimination and not
strong double-negation elimination, we take $B^*$ to be
$B$.
By Theorem \ref{theorem:pushback}, there is
a modus ponens proof of $B$ from substitution instances
of axioms.
If this proof contains any double negations that
do not occur in $B^*$, we simply erase them.  This
erasure takes a modus ponens step into another legal
modus ponens step.
Note that one cannot ``simply erase'' double negations
in a condensed-detachment proof;
but now we have a modus ponens proof, and double
negations {\em can} be erased in modus ponens proofs.
For axioms, the process transforms a substitution
instance of an axiom of L  into a substitution instance
of an axiom of L*.
Thus, we have a proof of $B^*$ from substitution
instances of axioms of L*
that contains no double negations except those that
already occur in $B^*$.
By Lemma \ref{lemma:axsub}, there exist condensed-detachment proofs of these substitution instances of L*
(from axioms of L*).  By hypothesis, the axioms of L*
have double-negation-free proofs in L.
We now construct the desired proof as follows.  First
write the double-negation-free proofs of the
axioms of L*.  Then write proofs of the substitution
instances of axioms of L* that are required.
These actions provide proofs of all the substitution
instances of axioms of L*, from L rather than from L*.
Now write the proof of $B^*$ from those substitution
instances.   We have the desired proof.  The
only double negations it contains are those contained
in $B^*$.
That completes the proof of the theorem.
  
Especially in view of the discussion focusing on the Frege axiom system, a natural question arises concerning its use as hypothesis.
In particular, if the theorem to be proved is itself free of double negation, must there exist a double-negation-free proof of it with the Frege system as hypothesis?
After all, that system contains two members exhibiting double negation.
Because we have in hand a proof that deduces from the Frege system the featured {\L}ukasiewicz axiom system such that the proof is free of double negation, such a proof must exist.
On the other hand, the following closely related question remains open and offers the researcher a most challenging problem to consider.
The question focuses on a condition stronger than strong double negation.
In particular, for two-valued sentential calculus, if the conclusion to be proved contains individual formulas that exhibit double negation, must there always exist a proof, say from the {\L}ukasiewicz three-axiom system, none of whose deduced steps exhibit double negation other than those formulas in the conclusion?
For example, we have a derivation of the Frege system from the {\L}ukasiewicz system such that exactly two of its deduced steps exhibit double negation, just the two members $i(n(n(x)),x)$ and $i(x,n(n(x)))$.

\begin{theorem} [Strong d-completeness]
\label{theorem:dcomplete2}
Suppose L is a logic that admits strong double-negation elimination.
If L proves $A$ using modus ponens and substitution,
without
using double negations except those that already occur
as subformulas of $A$, then
L proves $A$ using condensed detachment, without using
double negations except those that already occur as
subformulas of $A$.
\end{theorem}

\noindent{\em Proof\/}:  Suppose L proves $A$ using
modus ponens and substitution.   Then by Theorem
\ref{theorem:dcomplete},
there is a condensed-detachment proof of $A$ (possibly
using new double negations).  By strong double-negation
elimination, there is a condensed-detachment proof of
$A$ in L, using only double negations that already
occur as subformulas of $A$.

\section{{\L}ukasiewicz's System L1--L3}

As mentioned in Section 3, {\L}ukasiewicz's
system L has the following axioms.
\begin{eqnarray*}
\mbox{L1}& \qquad i(i(x,y),i(i(y,z),i(x,z))) \\
\mbox{L2}& \qquad i(i(n(x),x),x) \\
\mbox{L3}& \qquad i(x,i(n(x),y))
\end{eqnarray*}

\begin{lemma}  From L1--L3, one can find double-negation-free proofs of formulas D1--D3.
\end{lemma}
\noindent{\em Proof\/}.
Formula D1 is $i(x,x)$.  The following is a two-line
proof produced by OTTER.

\begin{tabbing}
31 \qquad \=[L1,L3] \qquad \=$i(i(i(n(x),y),z),i(x,z))$
\\
54 \>[31,L2] \>$i(x,x)$ \\
\end{tabbing}

Formula D2 is proved by first proving some auxiliary
formulas D4 and D5.
\begin{eqnarray*}
\mbox{D4} & \qquad i(i(x,i(x,y)),i(x,y)) \\
\mbox{D5} & \qquad i(i(x,y),i(n(y),n(x)))
\end{eqnarray*}

The following is an OTTER proof of D4 from L1--L3.
\begin{tabbing}
30 \qquad \= [L3,L2]\qquad \=$i(n(i(i(n(x),x),x)),y)$
\\
31 \>[L1,L1]\>$i(i(i(i(x,y),i(z,y)),u),i(i(z,x),u)) $
\\
32 \>[L1,30] \>$i(i(x,y),i(n(i(i(n(z),z),z)),y)) $ \\
33 \>[L1,L3] \>$i(i(i(n(x),y),z),i(x,z)) $ \\
34 \>[L1,L2] \>$i(i(x,y),i(i(n(x),x),y))  $ \\
35 \>[33,L2] \>$i(x,x) $ \\
36 \>[32,33] \>$i(x,i(n(i(i(n(y),y),y)),z))  $ \\
37 \>[31,31] \>$i(i(x,i(y,z)),i(i(u,y),i(x,i(u,z))))  $
\\
38 \>[31,34] \>$i(i(x,y),i(i(n(i(y,z)),i(y,z)),i(x,z)))
 $ \\
39 \>[38,37]
\>$i(i(x,i(n(i(y,z)),i(y,z))),i(i(u,y),i(x,i(u,z)))) $
\\
40 \>[39,36] \>$i(i(x,i(n(y),y)),i(z,i(x,y)))  $ \\
41 \>[40,31] \>$i(i(n(x),y),i(z,i(i(y,x),x))) $ \\
42 \>[41,39] \>$i(i(x,i(y,z)),i(i(n(z),y),i(x,z)))   $
\\
43 \>[42,37]
\>$i(i(x,i(n(y),z)),i(i(u,i(z,y)),i(x,i(u,y))))  $ \\
44 \>[42,L3] \>$i(i(n(x),n(y)),i(y,x)))  $ \\
45 \>[44,33] \>$i(x,i(y,x))  $ \\
46 \>[43,45] \>$i(i(x,i(y,z)),i(y,i(x,z)))  $ \\
47 \>[46,L3] \>$i(n(x),i(x,y))  $ \\
49 \>[47,42] \>$i(i(n(x),y),i(n(y),x))   $ \\
50 \>[49,43] \>$i(i(x,i(y,z)),i(i(n(y),z),i(x,z)))  $
\\
51 \>[50,35] \>$i(i(n(x),y),i(i(x,y),y)) $ \\
53 \>[51,47] \>$i(i(x,i(x,y)),i(x,y))  $ \\
\end{tabbing}

The following is an OTTER proof of D5 from L1--L3.
\begin{tabbing}
40 \qquad \=[L1,L1] \qquad
\=$i(i(i(i(x,y),i(z,y)),u),i(i(z,x),u))$ \\
41 \>[L1,L2] \>$i(i(x,y),i(i(n(x),x),y))$ \\
43 \>[L1,L3] \>$i(i(i(n(x),y),z),i(x,z))$ \\
44 \>[L3,L2] \>$i(n(i(i(n(x),x),x)),y)$ \\
46 \>[40,40] \>$i(i(x,i(y,z)),i(i(u,y),i(x,i(u,z)))$ \\
48 \>[40,2] \>$i(i(x,y),i(i(i(x,z),u),i(i(y,z),u)))$ \\
50 \>[40,41]
\>$i(i(x,y),i(i(n(i(y,z)),i(y,z)),i(x,z)))$ \\
65 \>[L1,44] \>$i(i(x,y),i(n(i(i(n(z),z),z)),y))$ \\
72 \>[46,48]
\>$i(i(x,i(i(y,z),u)),i(i(y,v),i(x,i(i(v,z),u))))$ \\
75 \>[46,50]
\>$i(i(x,i(n(i(y,z)),i(y,z))),i(i(u,y),i(x,i(u,z))))$
\\
84 \>[43,65] \>$i(x,i(n(i(i(n(y),y),y)),z))$ \\
97 \>[75,84] \>$i(i(x,i(n(y),y)),i(z,i(x,y)))$ \\
109 \>[40,97] \>$i(i(n(x),y),i(z,i(i(y,x),x)))$ \\
121 \>[75,109] \>$i(i(x,i(y,z)),i(i(n(z),y),i(x,z)))$
\\
124 \>[L1,109]
\>$i(i(i(x,i(i(y,z),z)),u),i(i(n(z),y),u))$ \\
130 \>[46,121]
\>$i(i(x,i(n(y),z)),i(i(u,i(z,y)),i(x,i(u,y))))$ \\
137 \>[121,L3] \>$i(i(n(x),n(y)),i(y,x))$ \\
144 \>[124,L2] \>$i(i(n(x),y),i(i(y,x),x))$ \\
158 \>[43,137] \>$i(x,i(y,x))$ \\
188 \>[130,158] \>$i(i(x,i(y,z)),i(y,i(x,z)))$ \\
201 \>[L1,158] \>$i(i(i(x,y),z),i(y,z))$ \\
232 \>[188,144] \>$i(i(x,y),i(i(n(y),x),y))$ \\
262 \>[201,137] \>$i(n(x),i(x,y))$ \\
309 \>[72,232] \>$i(i(n(x),y),i(i(z,x),i(i(y,z),x)))$
\\
422 \>[121,262] \>$i(i(n(x),y),i(n(y),x))$ \\
636 \>[422,158] \>$i(n(i(x,n(y))),y)$ \\
1158 \>[309,636]
\>$i(i(x,i(y,n(z))),i(i(z,x),i(y,n(z))))$ \\
1627 \>[1158,L3] \>$i(i(x,y),i(n(y),n(x)))$ \\
\end{tabbing}

Now we are ready to prove D2.  This proof was found by
Dolph Ulrich,
without machine assistance.
\begin{tabbing}
44 \qquad \= [D4,L1]        \=$i(i(x,x),i(x,x))$ \\
45  \> [L1,D5]
\>$i(i(i(n(y),n(x)),z),i(i(x,y),z))$ \\
43  \> [45,44]       \>$i(i(x,x),i(n(x),n(x)))$ \\
\end{tabbing}

Finally, we are ready to prove D3.  The following proof
was found
by using a specially compiled version of OTTER. (The
difficulty
is that normal OTTER derives a more general conclusion,
which subsumes the
desired conclusion.)

\begin{tabbing}
45 \qquad \= [L3,L2] \qquad \=$i(n(i(i(n(x),x),x)),y)$
\\
46 \>[L1,L1] \>$i(i(i(i(x,y),i(z,y)),u),i(i(z,x),u))$
\\
47 \>[L1,45] \>$i(i(x,y),i(n(i(i(n(z),z),z)),y))$ \\
48 \>[L1,L3] \>$i(i(i(n(x),y),z),i(x,z))$ \\
49 \>[L1,L2] \>$i(i(x,y),i(i(n(x),x),y))$ \\
50 \>[48,L2] \>$i(x,x)$ \\
51 \>[50,L1] \>$i(i(x,y),i(x,y))$ \\
52 \>[49,51] \>$i(i(n(i(x,y)),i(x,y)),i(x,y))$ \\
53 \>[47,48] \>$i(x,i(n(i(i(n(y),y),y)),z))$ \\
54 \>[53,L1] \>$i(i(i(n(i(i(n(x),x),x)),y),z),i(u,z))$
\\
55 \>[46,46] \>$i(i(x,i(y,z)),i(i(u,y),i(x,i(u,z))))$
\\
56 \>[46,L1] \>$i(i(x,y),i(i(i(x,z),u),i(i(y,z),u)))$
\\
57 \>[54,52] \>$i(x,i(i(n(y),y),y))$ \\
58 \>[55,57] \>$i(i(x,i(n(y),y)),i(z,i(x,y)))$ \\
59 \>[58,46] \>$i(i(n(x),y),i(z,i(i(y,x),x)))$ \\
60 \>[58,L3] \>$i(x,i(y,y))$ \\
61 \>[59,58] \>$i(x,i(i(n(y),z),i(i(z,y),y)))$ \\
62 \>[59,60] \>$i(x,i(i(i(y,y),z),z))$ \\
63 \>[61,61] \>$i(i(n(x),y),i(i(y,x),x))$ \\
64 \>[63,55] \>$i(i(x,i(y,z)),i(i(n(z),y),i(x,z)))$ \\
65 \>[63,48] \>$i(x,i(i(y,x),x))$ \\
66 \>[65,55] \>$i(i(x,i(y,z)),i(z,i(x,z)))$ \\
67 \>[66,65] \>$i(x,i(x,x))$ \\
68 \>[66,62] \>$i(x,i(y,x))$ \\
69 \>[67,60] \>$i(i(x,i(y,y)),i(x,i(y,y)))$ \\
70 \>[68,67] \>$i(i(x,i(y,x)),i(x,i(y,x)))$ \\
71 \>[64,55]
\>$i(i(x,i(n(y),z)),i(i(u,i(z,y)),i(x,i(u,y))))$ \\
72 \>[56,55]
\>$i(i(x,i(i(y,z),u)),i(i(y,v),i(x,i(i(v,z),u))))$ \\
73 \>[71,68] \>$i(i(x,i(y,z)),i(y,i(x,z)))$ \\
74 \>[73,L1] \>$i(i(i(x,i(y,z)),u),i(i(y,i(x,z)),u))$
\\
75 \>[74,69] \>$i(i(x,i(y,x)),i(y,i(x,x)))$ \\
76 \>[75,L1] \>$i(i(x,x),i(i(y,x),i(y,x)))$ \\
77 \>[72,70]
\>$i(i(x,i(i(y,z),u)),i(i(y,y),i(x,i(i(y,z),u))))$ \\
78 \>[77,76] \>$i(i(x,x),i(i(y,y),i(i(x,y),i(x,y))))$
\\
\end{tabbing}
That completes the proof of the lemma.

\begin{theorem} {\L}ukasiewicz's system L1--L3  admits
strong double-negation elimination.
\end{theorem}

\noindent{\em Proof}.  We begin by calculating the
formulas L* for this system.  We obtain the following.
\begin{eqnarray*}
\mbox{L4}& \qquad i(i(x,n(x)),n(x)) \\
\mbox{L5}& \qquad i(n(x),i(x,y))
\end{eqnarray*}
By Theorem \ref{theorem:main} it suffices to verify
that there exist double-negation-free
proofs of L4, L5, and D1--D3.  We have already verified
D1--D3 above, so it remains only
to exhibit double-negation-free proofs of L4 and L5.
The following is an OTTER proof of L4.
\begin{tabbing}
28  \qquad     \=[L1,L1]  \qquad
\=$i(i(i(i(x,y),i(z,y)),u),i(i(z,x),u))$ \\
29   \>[L1,L2]  \>$i(i(x,y),i(i(n(x),x),y))$ \\
31 \>[L1,L3]   \>$i(i(i(n(x),y),z),i(x,z))$ \\
32 \>[L3,L2]   \>$i(n(i(i(n(x),x),x)),y)$ \\
34 \>[28,28]   \>$i(i(x,i(y,z)),i(i(u,y),i(x,i(u,z))))$
\\
40 \>[28,29]
\>$i(i(x,y),i(i(n(i(y,z)),i(y,z)),i(x,z)))$ \\
54 \>[31,L2]    \>$i(x,x)$ \\
58 \>[L1,32]    \>$i(i(x,y),i(n(i(i(n(z),z),z)),y))$ \\
71 \>[34,40]
\>$i(i(x,i(n(i(y,z)),i(y,z))),i(i(u,y),i(x,i(u,z))))$
\\
94 \>[31,58]   \>$i(x,i(n(i(i(n(y),y),y)),z))$ \\
107 \>[71,94]  \>$i(i(x,i(n(y),y)),i(z,i(x,y)))$ \\
118 \>[28,107] \>$i(i(n(x),y),i(z,i(i(y,x),x)))$ \\
128 \>[71,118] \>$i(i(x,i(y,z)),i(i(n(z),y),i(x,z)))$
\\
141 \>[34,128]
\>$i(i(x,i(n(y),z)),i(i(u,i(z,y)),i(x,i(u,y))))$ \\
155 \>[128,L3]  \>$i(i(n(x),n(y)),i(y,x))$ \\
201 \>[31,155] \>$i(x,i(y,x))$ \\
262 \>[141,201] \>$i(i(x,i(y,z)),i(y,i(x,z)))$ \\
330 \>[262,L3]   \>$i(n(x),i(x,y))$ \\
421 \>[128,330] \>$i(i(n(x),y),i(n(y),x))$ \\
558 \>[141,421] \>$i(i(x,i(y,z)),i(i(n(y),z),i(x,z)))$
\\
731 \>[558,54] \>$i(i(n(x),y),i(i(x,y),y)$ \\
1032 \>[731,54] \>$i(i(x,n(x)),n(x))$ \\
\end{tabbing}

The following is an OTTER proof of L5.

\begin{tabbing}
19 \qquad \=[L1,L1] \qquad
\=$i(i(i(i(x,y),i(z,y)),u),i(i(z,x),u))$\\
20 \>[L1,L2] \>$i(i(x,y),i(i(n(x),x),y))$\\
22 \>[L1,L3] \>$i(i(i(n(x),y),z),i(x,z))$\\
23 \>[L3,L2] \>$i(n(i(i(n(x),x),x)),y)$\\
25 \>[19,19] \>$i(i(x,i(y,z)),i(i(u,y),i(x,i(u,z))))$\\
31 \>[19,20]
\>$i(i(x,y),i(i(n(i(y,z)),i(y,z)),i(x,z)))$\\
77 \>[L1,23] \>$i(i(x,y),i(n(i(i(n(z),z),z)),y))$\\
207 \>[25,31]
\>$i(i(x,i(n(i(y,z)),i(y,z))),i(i(u,y),i(x,i(u,z))))$\\
234 \>[22,77] \>$i(x,i(n(i(i(n(y),y),y)),z))$\\
265 \>[207,234] \>$i(i(x,i(n(y),y)),i(z,i(x,y)))$\\
287 \>[19,265] \>$i(i(n(x),y),i(z,i(i(y,x),x)))$\\
297 \>[207,287]
\>$i(i(x,i(y,z)),i(i(n(z),y),i(x,z)))$\\
389 \>[297,L3] \>$i(i(n(x),n(y)),i(y,x))$\\
439 \>[22,389] \>$i(x,i(y,x))$\\
522 \>[L1,439] \>$i(i(i(x,y),z),i(y,z))$\\
590 \>[522,389] \>$i(n(x),i(x,y))$
\end{tabbing}
That completes the proof of the theorem.

\begin{corollary}  Let $T$ be any set of axioms for
(two-valued) propositional logic.
Suppose that there exist double-negation-free
condensed-detachment proofs of L1--L3 from $T$.
Then the preceding theorem is true with $T$ in place of
L1--L3.
\end{corollary}

\noindent{\em Proof\/}.  We must show that $T$ admits strong
double-negation elimination.
Let $A$ be provable from $T$, and let $A*$ be obtained
from $A$ by erasing some of the double
negations in $A$ (but all occurrences of any given
formula, if there are multiple occurrences of
the same doubly negated subformula). We must show that
$T$ proves $A*$ by a proof whose doubly negated
subformula occur in $A*$.  Since $T$ is an
axiomatization of two-valued logic, $A*$ is a
tautology and hence provable
from L1--L3. By the theorem, there exists a proof of
$A*$ from L1--L3 that
contains no double negations (except those occurring in
$A*$, if any).  Supplying the given proofs of
L1--L3 from $T$,  we construct a proof of $A*$ from $T$
that contains no double negations except
those occurring in $A*$ (if any).  That completes the
proof.

{\em Example}.  We can take $T$ to contain exactly one
formula, the single axiom M of Meredith.
M is double-negation free, and double-negation-free
proofs of L1--L3 from M have been found
using OTTER \cite{Wos2001}.  Therefore, the theorem is
true for the single axiom M.

\section{Infinite-Valued Logic}
{\L}ukasiewicz's  infinite-valued logic is a subsystem
of classical propositional logic that was studied in
the 1930s.
The logic is of interest partly because there exists a
natural semantics for it,
according to which propositions are assigned truth
values that are real (or rational) numbers between 0
and 1, with 1 being true and 0 being false.
{\L}ukasiewicz's axioms
A1--A4 are complete for this semantics, as was proved
(but apparently not published) by Wasjberg, and proved
again by
Chang \cite{chang}.   Axioms A1--A4 are formulated
by using implication $i(p,q)$ and negation $n(p)$ only.
The truth value of $p$ is denoted by $\| p \| $.
 Truth values are given by
$$ \| n(p) \| = 1 - \| p \|$$
$$ \| i(p,q) \| = \min (1 - \|p\| + \|q\|,1).$$
Axioms A1--A4 are as follows.%
\footnote{
A comparison with {\L}ukasiewicz's axioms L1--L3:
Axiom A2 is the same as L1, and L3 is provable from
A1--A4, but L2 is not provable from A1--A4.}

\begin{eqnarray*}
\mbox{A1} &\hskip 0.5in &i(x,i(y,x)) \\
\mbox{A2}& &i(i(x,y),i(i(y,z),i(x,z)))\\
\mbox{A3}& &i(i(i(x,y),y),i(i(y,x),x))  \\
\mbox{A4}& &i(i(n(x),n(y)),i(y,x))
\end{eqnarray*}

\noindent
The standard reference for infinite-valued logic is
\cite{RoseAndRosser}.

\begin{lemma}  A1--A4 prove formulas D1--D3 without
double negation.
\end{lemma}
\noindent{\em Proof\/}.   The following is an OTTER
proof of D1 from A1--A4.
\begin{tabbing}
24 \qquad \=[A2,A1] \qquad \=$ i(i(i(x,y),z),i(y,z))$\\
32 \>[24,A3] \>$ i(x,i(i(x,y),y))$\\
59 \>[A2,32] \>$i(i(i(i(x,y),y),z),i(x,z))$\\
113 \>[59,24] \>$i(x,i(y,y))$\\
118 \>[113,113] \>$i(x,x)$
\end{tabbing}

The following is an OTTER proof of D2 from A1--A4,
found
by using a specially compiled version of OTTER.

\begin{tabbing}
118 \qquad \=[A1,A1] \qquad\= $i(x,i(y,i(z,y)))$\\
119\> [A2,A2]\>$i(i(i(i(x,y),i(z,y)),u),i(i(z,x),u))$\\
121\> [A2,A1]\>$i(i(i(x,y),z),i(y,z))$\\
122\> [A2,A3]\>$i(i(i(i(x,y),y),z),i(i(i(y,x),x),z))$\\
126\> [A3,118]\> $i(i(i(x,i(y,x)),z),z)$\\
139\> [121,A4]\>$i(n(x),i(x,y))$\\
140\> [121,A3]\> $i(x,i(i(x,y),y))$\\
143\> [A2,139]\> $i(i(i(x,y),z),i(n(x),z))$\\
148\> [A2,140]\> $i(i(i(i(x,y),y),z),i(x,z))$\\
179\> [121,126]\> $i(x,x)$\\
185\> [140,179]\> $i(i(i(x,x),y),y)$\\
214\> [A3,185]\> $i(i(x,i(y,y)),i(y,y))$\\
235\> [122,119]\>
$i(i(i(i(x,y),i(z,y)),i(z,y)),i(i(x,z),i(x,y)))$\\
385\> [119,148]\> $i(i(x,i(y,z)),i(y,i(x,z)))$\\
395\> [148,A3]\> $i(x,i(i(y,x),x))$\\
591\> [385,395]\> $i(i(x,y),i(y,y))$\\
598\> [591,591]\> $i(i(x,x),i(x,x))$\\
628\> [591,118]\> $i(i(x,i(y,x)),i(x,i(y,x)))$\\
645\> [143,598]\> $i(n(x),i(x,x))$\\
646\> [121,628]\> $i(i(x,y),i(y,i(x,y)))$\\
647\> [646,645]\> $i(i(x,x),i(n(x),i(x,x)))$\\
648\> [119,214]\> $i(i(x,y),i(x,x))$\\
648\> [A1,648]\> $i(x,i(i(y,z),i(y,y)))$\\
650\> [235,649]\> $i(i(x,i(y,z)),i(x,i(y,y)))$\\
651\> [650,647]\> $i(i(x,x),i(n(x),n(x)))$\\
\end{tabbing}

The following is a proof of D3, found
by using a specially compiled version of OTTER.

\begin{tabbing}
30 \qquad \= [A1,A1] \qquad \=$i(x,i(y,i(z,y)))$ \\
31 \>[A2,A2] \>$  i(i(i(i(x,y),i(z,y)),u),i(i(z,x),u))$
\\
32 \>[A2,A1] \>$  i(i(i(x,y),z),i(y,z))$ \\
33 \>[31,31] \>$  i(i(x,i(y,z)),i(i(u,y),i(x,i(u,z))))$
\\
34 \>[31,L2] \>$  i(i(x,y),i(i(i(x,z),u),i(i(y,z),u)))$
\\
35 \>[32,31] \>$  i(i(x,y),i(z,i(x,z)))$ \\
36 \>[32,A4] \>$  i(n(x),i(x,y))$ \\
37 \>[32,A1] \>$  i(x,i(y,i(z,x)))$ \\
38 \>[34,33] \>$
i(i(x,i(i(y,z),u)),i(i(y,v),i(x,i(i(v,z),u))))$ \\
39 \>[36,35] \>$  i(x,i(n(y),x))$ \\
40 \>[39,30] \>$  i(n(x),i(y,i(z,i(u,z))))$ \\
41 \>[A3,32] \>$  i(x,i(i(x,y),y))$ \\
42 \>[41,33] \>$  i(i(x,i(y,z)),i(y,i(x,z)))$ \\
43 \>[42,A2] \>$  i(i(i(x,i(y,z)),u),i(i(y,i(x,z)),u))$
\\
44 \>[42,A1] \>$  i(x,i(y,y))$ \\
45 \>[44,37] \>$  i(x,i(y,i(z,i(u,u))))$ \\
46 \>[44,A3] \>$  i(i(i(x,x),y),y)$ \\
47 \>[46,A3] \>$ i(i(x,i(y,y)),i(y,y))$ \\
48 \>[47,45] \>$ i(i(x,i(y,y)),i(x,i(y,y)))$ \\
49 \>[47,40] \>$  i(i(x,i(y,x)),i(x,i(y,x)))$ \\
50 \>[48,43] \>$  i(i(x,i(y,x)),i(y,i(x,x)))$ \\
51 \>[49,38] \>$
i(i(x,i(i(y,z),u)),i(i(y,y),i(x,i(i(y,z),u))))$ \\
52 \>[50,A2] \>$  i(i(x,x),i(i(y,x),i(y,x)))$ \\
53 \>[52,51] \>$  i(i(x,x),i(i(y,y),i(i(x,y),i(x,y))))$
\\
\end{tabbing}

\begin{theorem} The system of ``infinite-valued logic''
A1--A4  admits strong double-negation elimination.
\end{theorem}

\noindent{\em Proof}.  We begin by calculating the
formulas L* for this system.  The only
axiom containing negations is A4, but there are three
possible replacements, so we get
three new axioms A6--A8 as follows.%
\footnote{The name A5 is already in use for another
formula, originally used as an axiom
along with A1--A4, but later shown to be provable from
A1--A4.}
\begin{eqnarray*}
\mbox{A6}& \qquad i(i(x,y),i(n(y),n(x))) \\
\mbox{A7}& \qquad i(i(n(x),y),i(n(y),x))  \\
\mbox{A8}& \qquad i(i(x,n(y)),i(y,n(x)))
\end{eqnarray*}
By Theorem \ref{theorem:main} it suffices to verify
that there exist double-negation-free
proofs of A6, A7, A8, and D1--D3.  We have already
verified D1--D3 above, so it remains only
to produce double-negation-free proofs of A6--A8.

The following is an OTTER proof of A6.
\begin{tabbing}
81 \qquad \=[A1,A1]\qquad \=$ i(x,i(y,i(z,y)))$\\
82 \>[A2,A2] \>$i(i(i(i(x,y),i(z,y)),u),i(i(z,x),u))$\\
84 \>[A2,A1] \>$i(i(i(x,y),z),i(y,z))$\\
87 \>[A2,A4] \>$i(i(i(x,y),z),i(i(n(y),n(x)),z))$\\
89 \>[A3,81] \>$i(i(i(x,i(y,x)),z),z)$\\
92 \>[82,82] \>$i(i(x,i(y,z)),i(i(u,y),i(x,i(u,z))))$\\
95 \>[82,A2] \>$i(i(x,y),i(i(i(x,z),u),i(i(y,z),u)))$\\
99 \>[84,A4] \>$i(n(x),i(x,y))$\\
100 \>[84,A3] \>$i(x,i(i(x,y),y))$\\
112 \>[87,89] \>$i(i(n(x),n(i(y,i(z,y)))),x)$\\
148 \>[95,99] \>$i(i(i(n(x),y),z),i(i(i(x,u),y),z)))$\\
149 \>[92,99] \>$i(i(x,y),i(n(y),i(x,z)))$\\
154 \>[92,100] \>$i(i(x,i(y,z)),i(y,i(x,z)))$\\
291 \>[154,95]
\>$i(i(i(x,y),z),i(i(x,u),i(i(u,y),z)))$\\
296 \>[154,A2] \>$i(i(x,y),i(i(z,x),i(z,y)))$\\
450 \>[92,296]
\>$i(i(x,i(y,z)),i(i(z,u),i(x,i(y,u))))$\\
566 \>[450,149]\>$
i(i(i(x,y),z),i(i(x,u),i(n(u),z)))$\\
791 \>[566,112] \>$i(i(n(x),y),i(n(y),x))$\\
998 \>[148,791] \>$i(i(i(x,y),z),i(n(z),x))$\\
1109 \>[291,998]\>$
i(i(i(x,y),z),i(i(z,u),i(n(u),x)))$\\
1186 \>[1109,112]\>$ i(i(x,y),i(n(y),n(x)))$
\end{tabbing}

The following is an OTTER proof of A7.
\begin{tabbing}
81 \qquad \=[A1,A1] \qquad \=$i(x,i(y,i(z,y)))$\\
82 \>[A2,A2] \>$i(i(i(i(x,y),i(z,y)),u),i(i(z,x),u))$\\
84 \>[A2,A1] \>$i(i(i(x,y),z),i(y,z))$\\
87 \>[A2,A4] \>$i(i(i(x,y),z),i(i(n(y),n(x)),z))$\\
89 \>[A3,81] \>$i(i(i(x,i(y,x)),z),z)$\\
92 \>[82,82] \>$i(i(x,i(y,z)),i(i(u,y),i(x,i(u,z))))$\\
99 \>[84,A4] \>$i(n(x),i(x,y))$\\
100 \>[84,A3] \>$i(x,i(i(x,y),y))$\\
112 \>[87,89] \>$i(i(n(x),n(i(y,i(z,y)))),x)$\\
149 \>[92,99] \>$i(i(x,y),i(n(y),i(x,z)))$\\
154 \>[92,100] \>$i(i(x,i(y,z)),i(y,i(x,z)))$\\
296 \>[154,A2] \>$i(i(x,y),i(i(z,x),i(z,y)))$\\
450 \>[92,296]
\>$i(i(x,i(y,z)),i(i(z,u),i(x,i(y,u)))))$\\
566 \>[450,149]
\>$i(i(i(x,y),z),i(i(x,u),i(n(u),z)))$\\
791 \>[566,112] \>$i(i(n(x),y),i(n(y),x))$
\end{tabbing}

The following is an OTTER proof of A8.
\begin{tabbing}
81 \qquad \=[A1,A1] \qquad \=$i(x,i(y,i(z,y)))$\\
82 \>[A2,A2] \>$i(i(i(i(x,y),i(z,y)),u),i(i(z,x),u))$\\
84 \>[A2,A1] \>$i(i(i(x,y),z),i(y,z))$\\
87 \>[A2,A4] \>$i(i(i(x,y),z),i(i(n(y),n(x)),z))$\\
89 \>[A3,81] \>$i(i(i(x,i(y,x)),z),z)$\\
92 \>[82,82] \>$i(i(x,i(y,z)),i(i(u,y),i(x,i(u,z))))$\\
95 \>[82,A2] \>$i(i(x,y),i(i(i(x,z),u),i(i(y,z),u)))$\\
99 \>[84,A4] \>$i(n(x),i(x,y))$\\
100 \>[84,A3] \>$i(x,i(i(x,y),y))$\\
112 \>[87,89] \>$i(i(n(x),n(i(y,i(z,y)))),x)$\\
148 \>[95,99] \>$i(i(i(n(x),y),z),i(i(i(x,u),y),z))$\\
149 \>[92,99] \>$i(i(x,y),i(n(y),i(x,z)))$\\
154 \>[92,100] \>$i(i(x,i(y,z)),i(y,i(x,z)))$\\
291 \>[154,95]
\>$i(i(i(x,y),z),i(i(x,u),i(i(u,y),z)))$\\
296 \>[154,A2] \>$i(i(x,y),i(i(z,x),i(z,y)))$\\
450 \>[92,296]
\>$i(i(x,i(y,z)),i(i(z,u),i(x,i(y,u))))$\\
554 \>[82,450]
\>$i(i(x,y),i(i(z,u),i(i(y,z),i(x,u))))$\\
566 \>[450,149]
\>$i(i(i(x,y),z),i(i(x,u),i(n(u),z)))$\\
791 \>[566,112] \>$i(i(n(x),y),i(n(y),x))$\\
998 \>[148,791] \>$i(i(i(x,y),z),i(n(z),x))$\\
1109 \>[291,998]
\>$i(i(i(x,y),z),i(i(z,u),i(n(u),x)))$\\
1186 \>[1109,112] \>$i(i(x,y),i(n(y),n(x)))$\\
1207 \>[554,1186]
\>$i(i(x,y),i(i(i(n(z),n(u)),x),i(i(u,z),y)))$\\
1409 \>[1207,A4]
\>$i(i(i(n(x),n(y)),i(n(z),n(u))),i(i(y,x),i(u,z)))$\\
1561 \>[1409,791] \>$i(i(x,n(y)),i(y,n(x)))$
\end{tabbing}
This completes the proof of the theorem.

\section{An Intriguing Example}
One of the motivations for this work was the existence
of a formula that is double-negation free and
provable from A1--A4  but for
which Wos had been unable to find a
double-negation-free proof.   The formula
in question is

\begin{eqnarray*}
\mbox{DN1}& i(i(n(x),n(i(i(n(y),n(z)),n(z)))),\\
\mbox{   }& n(i(i(n(i(n(x),y)),n(i(n(x),z))),n(i(n(x),z))))).
\end{eqnarray*}

\noindent
Wos provided a proof of 45 condensed-detachment
steps of this theorem,
16 of whose lines involved a double negation.  Beeson
used this proof as input to a
computer program implementing the algorithms implicit
in the proof of our main
theorem.  The output of this program was a
double-negation-free proof by modus ponens
of the example,  from substitution instances of A1--A4.
The proof's length and size were surprising.
It was 796 lines, and
many of its lines involved thousands
of symbols.   The input proof takes about 3.5
kilobytes,  the output proof about
200 kilobytes.  Now we know what the {\em condensed}
means in ``condensed detachment''!
The expansion in size is due to making the
substitutions introduced by condensed detachment
explicit.  The expansion in length is due to
duplications of multiply referenced lines, which must
be
done before the substitutions are ``pushed upward'' in
the proof.  In other words, one line of
the proof can be referenced several times, and when the
proof is converted to tree form,
each reference will require a separate copy of the
referenced line.  This 796-line proof,
considered as a tree,  has substitution instances of
the axioms at the leaves.   After obtaining this
proof, we could have continued with the algorithm,
providing proofs of the substitution instances of
the axioms.   That approach would have substantially increased
the length.   Instead, McCune
put the lines of the 796-line proof into an OTTER input
file as ``hints'' \cite{veroff-hints}, and OTTER
produced a 27-line double-negation-free
condensed-detachment proof of DN1 from A1--A4 and
A6--A8.
This run generates some 6,000 formulas and takes between
one-half and two hours, depending on what machine is used.
If the lines of this proof, together with the proofs of
A6--A8, are supplied as resonators \cite{wos-resonance} in a new
input file,  OTTER can then find a 37-step proof of DN1
from A1--A4. 

\section{D-Completeness of Intuitionistic Logic}

Let H be the following formulation of intuitionistic
propositional calculus
in terms of implication and negation,
denoted by $i$ and $n$.%
\footnote{These axioms can be found in Appendix I of \cite{prior}, as the {\L}ukasiewicz 2-basis in 12.1
plus the two axioms labeled (4) of 3.2, as specified in 12.5.  According to \cite{prior}, if we also add (2) and (3) of 3.2, we get the full
intuitionistic propositional calculus; but (2) and (3) of 3.2 concern disjunction and conjunction.  If they are omitted,
the four axioms listed form a 4-basis for the implication-negation fragment.  This will be proved in Corollary
\ref{corollary:basis}.}
\begin{eqnarray*}
\mbox{H1}& \qquad i(x,i(y,x)) \\
\mbox{H2}& \qquad i(i(x,i(y,z)),i(i(x,y),i(x,z))) \\
\mbox{H3}& \qquad i(i(x, n(x)), n(x)) \\
\mbox{H4}& \qquad i(x, i(n(x), y))
\end{eqnarray*}
The inference rules of H are modus ponens and
substitution.   It is also possible
to consider H1--H4 with condensed detachment.   These two systems have the same
theorems, as will be shown in detail below.

We note that
H does not satisfy strong double elimination.  Substituting $n(y)$ for $x$ in 
axiom H3 produces $i(i(n(y),n(n(y))),n(n(y)))$.  Cancelling the double negations
produces  $i(i(n(y),y),y)$, which is not provable in 
intuitionistic logic.  This same example demonstrates directly that H does not 
satisfy the hypothesis of Theorem \ref{theorem:main}.  Nevertheless, and perhaps 
surprising, H does satisfy double negation elimination---but we will need a different 
proof to show that.  

\begin{lemma} \label{lemma:HD1--D3} D1--D3 have
double-negation-free condensed-detachment proofs from
H1--H4.
\end{lemma}

\noindent{\em Proof\/}:
The following is a double-negation-free
condensed-detachment proof of D1 from H1--H4
(found by hand):

\begin{tabbing}
5 \qquad \=[H2,H1]\qquad \= $i(i(x,y),i(x,x))$\\
6 \>[5,H1]\> $i(x,x)$
\end{tabbing}

D2 is  $i(i(x,x),i(n(x),n(x)))$.   The following is a
double-negation-free condensed-detachment proof of D2
from H1--H4,
found by using a specially compiled version of OTTER.
Curiously, H3 is not used.

\begin{tabbing}
94 \qquad \=[H1,H1]\qquad \= $i(x,i(y,i(z,y)))$\\
95
\>[H2,H2]\>$i(i(i(x,i(y,z)),i(x,y)),i(i(x,i(y,z)),i(x,z
)))$\\
97 \>[H2,H1] \>$i(i(x,y),i(x,x))$\\
100 \>[H2,H4]\>$i(i(x,n(x)),i(x,y))$\\
107 \>[H1,94] \>$i(x,i(y,i(z,i(u,z))))$\\
111 \>[95,94] \>$i(i(x,i(i(y,x),z)),i(x,z))$\\
113 \>[95,97] \>$i(i(x,i(x,y)),i(x,y))$\\
116 \>[H1,97] \>$i(x,i(i(y,z),i(y,y)))$\\
213 \>[H2,116] \>$i(i(x,i(y,z)),i(x,i(y,y)))$\\
220 \>[H1,100] \>$i(x,i(i(y,n(y)),i(y,z)))$\\
753 \>[95,113] \>$i(i(x,i(x,x)),i(x,x))$\\
783 \>[753,116] \>$i(i(x,x),i(x,x))$\\
785 \>[753,107] \>$i(i(x,i(y,x)),i(x,i(y,x)))$\\
833 \>[783,213]\>$i(i(x,i(y,y)),i(x,i(y,y)))$\\
903 \>[785,94] \>$i(i(x,y),i(y,i(x,y)))$\\
1541 \>[111,220] \>$i(n(x),i(x,y))$\\
1563 \>[833,1541] \>$i(n(x),i(x,x))$\\
1605 \>[903,1563] \>$i(i(x,x),i(n(x),i(x,x)))$\\
1706 \>[213,1605] \>$i(i(x,x),i(n(x),n(x)))$
\end{tabbing}

The following is a double-negation-free
condensed-detachment proof of D3 from H1--H4.
Again H3 is not used.

\begin{tabbing}
177 \qquad\= [H1,H1] \qquad\= $i(x,i(y,i(z,y)))$\\
178
\>[H2,H2]\>$i(i(i(x,i(y,z)),i(x,y)),i(i(x,i(y,z)),i(x,z
)))$\\
179 \>[H1,H2]
\>$i(x,i(i(y,i(z,u)),i(i(y,z),i(y,u))))$\\
180 \>[H2,H1] \>$i(i(x,y),i(x,x))$\\
194 \>[178,177] \>$i(i(x,i(i(y,x),z)),i(x,z))$\\
196 \>[178,180] \>$i(i(x,i(x,y)),i(x,y))$\\
273 \>[2,194] \>$i(x,i(i(y,i(i(z,y),u)),i(y,u)))$\\
275 \>[194,179] \>$i(i(x,y),i(i(z,x),i(z,y)))$\\
310 \>[3,275]
\>$i(i(i(x,y),i(z,x)),i(i(x,y),i(z,y)))$\\
351 \>[194,273] \>$i(i(i(x,y),z),i(y,z))$\\
442 \>[351,310]\>$i(i(x,y),i(i(y,z),i(x,z)))$\\
655 \>[442,442]
\>$i(i(i(i(x,y),i(z,y)),u),i(i(z,x),u))$\\
1010 \>[655,655]
\>$i(i(x,i(y,z)),i(i(u,y),i(x,i(u,z))))$\\
1036 \>[655,442]
\>$i(i(x,y),i(i(i(x,z),u),i(i(y,z),u)))$\\
1355 \>[1010,1036]
\>$i(i(x,i(i(y,z),u)),i(i(y,v),i(x,i(i(v,z),u))))$\\
2170 \>[196,275] \>$i(i(x,x),i(x,x))$\\
2188 \>[275,2170] \>$i(i(x,i(y,y)),i(x,i(y,y)))$\\
2211 \>[2170,177] \>$i(i(x,i(y,x)),i(x,i(y,x)))$\\
2335 \>[2188,275] \>$i(i(x,x),i(i(y,x),i(y,x)))$\\
2404 \>[2211,1355]
\>$i(i(x,i(i(y,z),u)),i(i(y,y),i(x,i(i(y,z),u))))$\\
2537 \>[2404,2335]
\>$i(i(x,x),i(i(y,y),i(i(x,y),i(x,y))))$
\end{tabbing}

\begin{theorem} \label{theorem:dcompleteH} The same
theorems are provable from H1--H4
by using condensed detachment as
the sole rule of inference as when we use modus ponens
and substitution as rules of inference.   Moreover,
if $b$ is provable without double negation by modus
ponens from substitution instances of axioms, then
there is a double-negation-free condensed-detachment
proof of $b$.
\end{theorem}

\noindent{\em Remark}.  The present proof gives no
assurance that a general H proof,
using substitution arbitrarily and not just in axioms,
can be converted to a condensed-detachment
proof without introducing additional double negations.
That in general it can be so converted will follow from
Theorem \ref{theorem:H1}.

\noindent{\em Proof\/}: The first claim is an immediate
consequence of
 Theorem \ref{theorem:dcomplete} and Lemma
\ref{lemma:HD1--D3}.  To prove the second claim,
suppose $b$ has a double-negation-free modus ponens
proof from substitution instances of axioms.
By Lemma \ref{lemma:axsub}, we can supply
double-negation-free condensed-detachment proofs of the
substitution instances of axioms that are used in the
proof.  Adjoining these proofs, we obtain a
double-negation-free condensed-detachment proof of $b$
as required.

\section{H and Sequent Calculus}
Let G1 be the intuitionistic Gentzen calculus as given
by Kleene \cite{im}.
Let G be G1 (minus cut), restricted to implication
and negation; that is, formulas containing other
connectives are not allowed.
The rules of inference of G are the four rules
involving implication and negation,
plus the structural rules.\footnote{The theorems of G include all theorems of G1 that
involve only implication and negation, since by Gentzen's
cut-elimination theorem, any theorem has a cut-free proof,
and the formulas appearing in the proof are all subformulas
of the final sequent and hence involve only the connectives
that occur in the final sequent.}
The rules of G1 are listed
on pp. 442--443 of \cite{im}.  They will also
be given in the course of the proof of Lemma
\ref{lemma:GtoH}.
We shall use the notation $\Gamma \seq \Delta$ for a
sequent.
We remind the reader that what distinguishes
intuitionistic from classical sequent calculus is that
the consequent $\Delta$ in a sequent $\Gamma \seq
\Delta$ in the intuitionistic calculus is restricted
to contain at most one formula.%
\footnote{The translations given here can also be given
for {\L}ukasiewicz's logic L1--L3, but many
additional complications are introduced by the
necessity of translating a sequent containing more than
one formula on the right, and in view of the simpler
proofs of double-negation elimination given above, we
treat
the Gentzen translation only for intuitionistic logic.
Note that we used OTTER only for the H-proofs
of D2 and D3;  but if we treat L this way instead of H,
we need OTTER for  twenty-one additional lemmas.}

We give a translation of H into G: If $A$ is a formula
of H, then $A^0$ is a formula of G,  obtained by the following
rules.
\begin{eqnarray*}
i(a,b)^0 = a^0 \implies b^0 \\
n(a)^0 = \neg a^0
\end{eqnarray*}
Of course, when $a$ is a proposition letter (variable),
then $a^0$ is just $a$.
If $\Gamma = A_0,\ldots,A_n$ is a list of formulas of
L, then $\Gamma^0$ is the list
$A_0^0,\ldots,A_n^0$.

We translate G into H in the following manner.  First
we  assign to each formula $A$ of G a corresponding
formula $A^{\prime}$ of H, given by
\begin{eqnarray*}
(A \implies B)^\prime = i(A^\prime, B^\prime) \\
(\neg A)^\prime = n(A^\prime) ,
\end{eqnarray*}
where again $A^\prime = A$ for proposition letters $A$.
We need to define $\Gamma^\prime$ also, where $\Gamma$
is a list of formulas; since we are treating
only the intuitionistic calculus, we need this
definition only for
lists occurring on the left of $\seq$.
If $\Gamma = A_1,\ldots,A_n$ is a list of formulas
occurring on the left of $\seq$, then $\Gamma^\prime$
is
$A_1^\prime,\ldots,A_n^\prime$.%
\footnote{
A similar translation has been given in
\cite{prijatelj} in connection with
{\L}ukasiewicz's multivalued logics (which include the
infinite-valued logic discussed in Section 7
of this paper).  It is the obvious translation of
Gentzen calculus into the implication-and-negation
fragment of propositional calculus.  We cannot appeal
to any of the results of \cite{prijatelj} because we
are dealing with different logics, and besides we need
to pay attention to double negations.
}

These two translations are inverse.

\begin{lemma}
\label{lemma:inverse}
Let $A$ be a formula of H.  Then ${A^0}^\prime = A$.
\end{lemma}

\noindent{\em Proof}.  By induction on the complexity
of $A$.  If $A$ is a variable, then
$A^0 = A$ and ${A^0}^\prime = A$.  We have
\begin{eqnarray*}
{i(x,y)^0}^\prime &=& (x^0 \implies y^0)^\prime \\
  &=& i({x^0}^\prime, {y^0}^\prime) \\
  &=& i(x,y) ,
\end{eqnarray*}
and we have
\begin{eqnarray*}
{n(x)^0}^\prime &=& (\neg(x^0))^\prime \\
  &=& n({x^0}^\prime) \\
  &=& n(x) .
\end{eqnarray*}

Henceforth we simplify our notation by using lower-case
letters for formulas of H
and upper-case letters for formulas of G.  Then we can
write $a$ instead of $A^\prime$
and $A$ instead of $a^0$.  By the preceding lemma,
this convenient notation presents no ambiguity.
Thus, for example, $(A \implies B)^\prime$
is $i(a,b)$.  Greek letters are
used for lists of formulas.

The following lemma gives several variations of the
deduction theorem for H.

\begin{lemma}[Deduction theorem for H]  (i) If H proves
$a$ from assumptions $\delta, b$, then
$i(b,a)$ is a theorem proved in H from assumptions
$\delta$, provided the assumptions contain
only constant proposition letters.

(ii)  If $a$ is provable from assumptions $\delta, b$
by condensed detachment from H1--H4,
then $i(b,a)$ is derivable by condensed detachment from
$\delta$, provided the assumptions contain
only constant proposition letters.

(iii) If there exists a  proof of $a$ by modus ponens from
$\delta,b$ and substitution instances of  H1--H4,
then there exists a proof of
$i(a,b)$ by modus ponens from $\delta$ and substitution
instances of  H1--H4.

(iv) In part (i), if the given proof of $a$ has no
double negations,
then the proof of $i(b,a)$ from $\delta$ has no  double
negations.

(v) In part (iii), if the given proof of $a$ has no
double negations,
then the proof of $i(b,a)$ from $\delta$ has no  double
negations.

\end{lemma}

\noindent{\em Remarks\/}: We do not prove a claim about
double negations
for condensed-detachment proofs, only for modus ponens
proofs.  That is, for condensed-detachment proofs,  
there is no part (vi)
analogous to parts (iv) and (v).
The reason for the
restriction to constant assumptions in (i) and (ii) is
the following.
From the assumption $i(n(n(x)),x)$,  we can derive any
theorem of classical logic, for instance
$i(n(n(a)),a)$,
by substitution or condensed detachment.  But we cannot
derive the proposition that the first of these implies
the second, $i(i(n(n(x)),x),i(n(n(a)),a))$.   Therefore the
deduction theorem is false without the restriction.
Proofs by modus ponens from substitution instances of
axioms do not suffer from this difficulty, which
is one reason they are so technically useful in this
paper.

\medskip
\noindent{\em Proof}.  First we show that (ii) follows
from (iii).  If we are given a 
condensed-detachment proof of $a$ from assumptions
$\delta,b$ using H1--H4,
we can find, by Theorem \ref{theorem:pushback}, a
modus ponens
proof of $a$ from $\delta, b$ and substitution
instances of H1--H4.
Applying (iii), we have a modus ponens proof of $i(b,a)$
from $\delta$ and substitution instances of H1--H4.
By Theorem \ref{theorem:dcompleteH}, this proof can be
converted to a condensed-detachment proof of $i(b,a)$
from  $\delta$, completing the derivation of (ii) from
(iii).

Next we show that (i) follows from (iii).
Suppose we are given a proof of $a$ from $\delta,b$ in
H.  By Theorem \ref{theorem:pushback},
we can find a modus ponens proof of $a$ from
assumptions $\delta, b$ and substitution instances
of H1--H4.   By (iii) we then can find a modus ponens
proof of $i(b,a)$ from $\delta$ and substitution
instances
of  H1--H4.  Adding one substitution step above each
such substitution instance, we have
a proof in $H$ of $i(b,a)$ from $\delta$.  That
completes the proof that (i) follows from (iii).

We now show that (v) implies (iv).   To do so requires
going over the preceding paragraph with attention to
double negations.
Suppose we are given a double-negation-free proof of
$a$ from $\delta,b$ in H.  By Theorem
\ref{theorem:pushback},
we can find a modus ponens proof of $a$ from
assumptions $\delta, b$ and substitution instances
of H1--H4, which is also double-negation free.   By
(v)
 we then can find a double-negation-free modus ponens
proof of $i(b,a)$ from $\delta$ and substitution
instances
of  H1--H4.  Adding one substitution step above each
such substitution instance, we have
a double-negation-free proof in $H$ of $i(b,a)$ from
$\delta$.  That completes the proof that (iv) follows
from (v).

Now we prove (iii) and (v) simultaneously by induction
on the number of steps in a pure modus ponens
proof of $a$ from $\delta$ and substitution instances
of H1--H4.

Base case:   $a$ is  $b$, or a member of $\delta$, or
a substitution instance of one of H1--H4.

If $a$ is a substitution instance of an axiom of
H1--H4, then by Lemma \ref{lemma:axsub} there
exists a condensed-detachment proof of $a$ from H1--H4
that contains only double negations already occurring
in $a$.

If $a$ is $b$, then we use the fact that $i(b,b)$ is a
theorem of H, provable without double negations (except
those occurring in $b$) by Lemma
\ref{lemma:ialphaalpha}.
Hence by Theorem \ref{theorem:pushback}, it is provable
by modus ponens from substitution instances of H1--H4.

If $a$ is a member of $\delta$, then consider the
formula $i(a,i(b,a))$, which is a substitution
instance of axiom H1.  We can deduce $i(b,a)$ by modus
ponens from this formula and $a$; adjoining this
step to a one-step proof of $a$ from $\delta$ ``by
assumption'', we have a proof of $i(b,a)$ from
$\delta$.

Turning to the induction step, suppose the last step in
the given proof infers $a$ from $i(p,a)$ and $p$.
By the induction hypothesis, we have proofs of $i(b,p)$
and $i(b,i(p,a))$ from $\delta$.  By axiom
H2 and modus ponens (which is a special case of
condensed detachment)
we have $i(i(b,p),i(b,a))$.  Applying modus ponens once
more, we have
$i(b,a)$ as desired.  Note that
no double negations are introduced.  That completes the
proof of the lemma.

We shall call a sequent $\Gamma \seq \Delta$
double-negation free if it
contains no double negation.

We shall refer to proofs by modus ponens
from substitution instances of H1--H4 as M-proofs  for
short.
Thus M-proofs use modus ponens only but can use
substitution instances
of axioms,  as opposed to H-proofs, which  can use
substitution anywhere as well as modus ponens.
We have already shown how to convert
condensed-detachment proofs to M-proofs
(in Theorem \ref{theorem:pushback}), and vice versa
(since every substitution
instance of the axioms is derivable by condensed
detachment).

\begin{lemma}
If the final sequent $\Gamma \seq \Theta$ of a G-proof
is double-negation free,  then the entire G-proof is
double-negation free.
\end{lemma}

\noindent{\em Proof}.  By the subformula property of
cut-free proofs:  Every formula
in the proof is a subformula of the final sequent.

\begin{lemma} \label{lemma:HtoGsound} The translation
from H to G is sound.  That is, if H proves $a$ from
assumptions $\delta$, then G proves the sequent $\Delta
\seq A$ (where $A$ is the translation $a^0$, and
$\Delta$ is $\delta^0$).
\end{lemma}

\noindent{\em Proof}.  We proceed by induction on the
length of proofs.  When the length is zero,
we must exhibit a proof in G of each of the axioms
H1--H4.  This is a routine exercise in
the Gentzen sequent calculus, which we omit.  For the
induction step,  suppose
we have proofs in H from assumptions $\delta $ of $a$
and $i(a,b)$.  Then by the induction hypothesis, we
have
proofs in G of $\Delta \seq A$ and $\Delta \seq A
\rightarrow B$.  It is another
exercise in Gentzen rules to produce a proof of $\Delta
\seq B$.  One solves this
exercise by first proving $$A \implies((A \implies B)
\implies B)$$
and then using the cut rule twice.  This completes the
proof of the lemma.

\begin{lemma} \label{lemma:GtoH}
(i) Suppose G proves the sequent $\Gamma \seq A$.
Then there is an M-proof of $a$ from assumptions
$\gamma$.  If G proves
$\Gamma \seq \>[]$,  where $[]$ is the empty list, then
there is an M-proof of $p$ from assumptions
$\gamma$, where $p$ is any formula of H.

(ii)  If any double negations occur in subformulas of
the given sequent $\Gamma \seq \Delta$
(where here $\Delta$ can be empty or not),
then a proof as in (i) can be found that contains no
double negations except those arising
from the H-translations of double-negated subformulas
of $\Gamma \seq \Delta$.

(iii)  If in part (i) the H-translation of the given
sequent $\Gamma \seq \Delta$
does not contain any double negations, then the M-proof
that is asserted to exist can also
be found without double negations.

\end{lemma}

\noindent
{\em Proof}.  We proceed by induction on the length of
proof of $\Gamma \seq A$ in G. 

Base case:
the sequent has the form $\Gamma, A \seq A$.  We must
show that $a$ is derivable in H from
premisses $\gamma, a$,  which is clear.  

Now for the
induction step.  We consider one case for
each rule of G.

Case 1,  the last inference in the G-proof is by rule
$\implies \seq$:
\medskip

\prooftree
    \Delta \seq A
    \qquad B, \Gamma \seq \Theta
\justifies
     A \implies B, \Delta, \Gamma \seq \Theta
\thickness=0.08em
\endprooftree
\medskip

By the induction hypothesis, we have an M-proof of $a$
from $\delta$,
and an M-proof of $\theta$ from $b$ and $\gamma$.
We must give an M-proof of $\theta$ from $i(a,b)$,
$\delta$,
and $\gamma$.

Applying modus ponens to $i(a,b)$ (which is $(A
\implies B)^\prime$)
and the given
proof of $a$ from $\delta$, we derive $b$.  Copying the
steps of the
proof of $\theta$ from assumptions $b, \gamma$  (but
changing the justification of the
step(s) $b$ from ``assumption'' to the line number
where $b$ has been derived), we
have derived $\theta$ from assumptions $(A \implies
B)^\prime$,$\delta, \gamma$,
completing the proof of case 1.   No double negations
are introduced by this step.

Case 2,  the last inference in the G-proof is by rule
$\seq \implies$:
\medskip

\prooftree
    A, \Gamma \seq B
\justifies
    \Gamma \seq  A \implies B
\thickness=0.08em
\endprooftree
\medskip

By the induction hypothesis, we have an M-proof from
H1--H4 of $b$ from $\gamma$ and $a$.
Applying the deduction theorem for H1--H4 with
M-proofs, we have an M-proof in H of $i(a,b)$ from
$\gamma$.
But $(A \implies B)^\prime = i(a,b)$, completing this
case.   Note that double negations
are not introduced by the deduction theorem if they are
not already present, by part (v) of
the deduction theorem.   (One sees why we must use
M-proofs instead of condensed detachment.)

Case 3,  the last inference in the G-proof introduces
negation on the right:
\medskip

\prooftree
    A, \Gamma \seq []
\justifies
    \Gamma \seq \neg A
\thickness=0.08em
\endprooftree
\medskip

By the induction hypothesis, there is an M-proof of
$n(a)$ from
$a$ and $\gamma$.
By the deduction theorem for H1--H4 with M-proofs,
there is a proof of $i(a,n(a))$ from
$\gamma$. Hence, it suffices to show that $n(a)$ is
derivable from $i(a, n(a))$.
This follows from a substitution instance of H3, which
is $i(i(x,n(x)),n(x))$,
substituting $a$ for $x$.

Case 4, the last inference in the G-proof introduces
negation on the left:
\medskip

\prooftree
\Gamma \seq A
\justifies
\neg A, \Gamma \seq []
\thickness=0.08em
\endprooftree
\medskip

By the induction hypothesis, we have an M-proof of $a$
from $\gamma$.
We must show that from $n(a)$ and $\gamma$, we can
deduce $b$ in L, where $b$ is any formula of $H$.
We have $i(a,i(n(a),b)$ as a substitution instance of
axiom H4.  Applying modus ponens twice,
we have the desired M-proof of $b$ from $\gamma$,
completing case 4.

Case 5,  the last inference is by contraction in the
antecedent:
\medskip

\prooftree
C,C, \Gamma \seq \Theta
\justifies
C,\Gamma \seq \Theta
\thickness=0.08em
\endprooftree
\medskip

By the induction hypothesis we have an M-proof of
$\theta$
from assumptions $c,c$,  which also qualifies as a
proof from assumptions $c$, so there is nothing
more to prove.

Case 6, the last inference is by thinning in the
antecedent:
\medskip

\prooftree
   \Gamma \seq \Theta
\justifies
   C, \Gamma \seq \Theta
\thickness=0.08em
\endprooftree
\medskip

By the induction hypothesis, we have an M-proof from
H1--H4 of $\theta$ from assumptions
$\Gamma^\prime$.  That counts as an M-proof from
assumptions $C, \gamma$ as well.
That completes case 6.

Case 7, the last inference is by interchange in the
antecedent.  This just means the order of formulas
in the assumption list has changed, so there is nothing
to prove.

That completes the proof of part (i) of the lemma.
Regarding parts (ii) and (iii): by the preceding lemma,
any double negations occurring anywhere in the G-proof
must occur in the final sequent.  No new
double negations are introduced in the translation to
H,  and all the theorems of H that we used
have been given double-negation-free
condensed-detachment proofs from H1--H4.  By Theorem
\ref{theorem:pushback},
they have double-negation-free M-proofs, too.
Although we may not have pointed it out in each case,
the argument given  produces
an M-proof in which any double negations arise from the
translations into H of doubly negated
subformulas of the final sequent.   In particular, if
the final sequent contains no double negations,
then the M-proof produced also contains no double
negations.

\begin{corollary} \label{corollary:basis} 
H is a basis for the implication-negation 
fragment of intuitionistic logic.  That is, every 
intuitionistically valid formula in this fragment is provable in H.
\end{corollary}

\noindent{\em Remark}.
In \cite{horn}, this lemma was proved
for a different axiomatization of
the implication-negation fragment of intuitionistic
calculus, so this corollary could also be proved
by demonstrating the equivalence of the two fragments
directly.

\noindent{\em Proof}. Suppose $A$ is an intuitionistically 
valid formula containing no connectives other than implication
and negation.  By Gentzen's cut-elimination theorem, there 
is a cut-free proof of the sequent $ [] \seq A$ (with empty 
antecedent).  By Lemma \ref{lemma:GtoH}, $A$ has an $M$-proof,
which in particular is a proof in H.  

\noindent{\em Remark}. The main idea of the corollary is that
by the subformula property of cut-free proofs, the cut-free
proof contains no connectives other than implication and negation.  

\begin{theorem} \label{theorem:H1}  Suppose H proves
$b$ from assumptions $\delta$ and neither $\delta$ nor
$b$ contains
double negation.  Then there is a condensed-detachment
proof of $b$ from H1--H4 and assumptions
$\delta$ that does not contain double negation.

More generally, if $\delta$ and $b$ are allowed to
contain double negation, then there is a condensed-detachment proof
of $b$ from H1--H4 and assumptions $\delta$ that
contains no new double negations.
That is, all doubly negated formulas occurring
in the proof are subformulas of $\delta$ or of $b$.
\end{theorem}

\noindent{\em Proof}.  Let $b^0 = B$ be the translation
of $b$ into G defined earlier.  Double negations in
$B$ arise only from double negations in $b$. Suppose
$b$ is provable in H from assumptions $\delta$.
By Theorem \ref{theorem:pushback}, there is an
M-proof of $b$ from $\delta$.   By Lemma
\ref{lemma:HtoGsound}, the
sequent $\Delta \seq B$ is provable in $G$.
Hence, by Gentzen's cut-elimination theorem, there is a
proof in G of $\Delta \seq b$.
By the previous lemma, there is an M-proof  of
${B}^\prime$ from assumptions
${\Delta^0}^\prime$ that contains no new double
negations.
But by Lemma \ref{lemma:inverse},  ${B}^\prime = b$ and
${\Delta}^\prime = \delta$.  Thus
we have an M-proof of $b$ from $\delta$.  By the
d-completeness of $H$,  Theorem
\ref{theorem:dcompleteH},
there is also a condensed-detachment proof of $b$ from
$\delta$.  The second part of Theorem
\ref{theorem:dcompleteH}
says that we can find a double-negation-free condensed-detachment proof of $b$ from $\delta$.  It is
important that we are working with M-proofs here, since
the second part of the D-completeness theorem,
about double negations, applies only to M-proofs.
That completes the proof.

\begin{theorem}  Suppose $A$ is provable from H1--H4
by using condensed detachment as the only rule of
inference.  Then $A$ has a proof from H1--H4 using
condensed detachment in which no doubly negated
formulas occur except those that already occur as
subformulas of $A$.
\end{theorem}

\noindent{\em Proof\/}.  Suppose $A$ is provable from
H1--H4 using condensed detachment.  Each
condensed-detachment step can be converted to three
steps by using substitution and modus ponens, so
$A$  is provable in H.  By the preceding theorem, $A$
has a condensed-detachment proof from H1--H4 in which
no doubly negated
formulas occur except those that already occur in $A$.
That completes the proof.

\noindent{\em Remark\/}.  Since the translation back
from Gentzen calculus produces
M-proofs, we do not need to appeal to d-completeness
for arbitrary H-proofs.  This is fortunate because we do
not
know a proof of d-completeness that avoids the possible
introduction of double negations, except when
restricted to M-proofs.

\begin{corollary}  Let $T$ be any set of axioms for
intuitionistic propositional logic.
Suppose that there exist condensed-detachment proofs of
H1--H4 from $T$ in which no double negations
occur (except those that occur in $T$, if any).  Then
the preceding theorem is true with $T$ in
place of H1--H4.
\end{corollary}

\noindent{\em Proof\/}.  Let $b$ be provable from $T$.
Then $b$ is provable
from H1--H4, since $T$ is a set of axioms for
intuitionistic logic.
 By the theorem, there is a proof of $b$ from H1--H4
that
contains no double negations (except those occurring in
$b$, if any).  Supplying the given proofs of
H1--H4 from $T$,  we construct a proof of $b$ from $T$
that contains no double negations except
those occurring in $T$ or in $b$ (if any).  That
completes the proof.

\section*{Acknowledgments}

We thank Kenneth Harris, Branden Fitelson, and Dolph
Ulrich
for their attention to early drafts of this paper.  In
addition, Ulrich
contributed part of the
proof of D1--D3 from L1--L3.

The work of RV was supported in part by National Science Foundation grant no. CDA-9503064.

The work of LW was supported by the Mathematical, Information, and Computational Sciences Division subprogram of the Office of Advanced Scientific Computing Research, U.S. Department of Energy, under Contract W-31-109-Eng-38.

\end{document}